\documentclass[a4paper,11pt]{article}
\usepackage[utf8]{inputenc}

\usepackage{framed}
\usepackage{mdframed}

\usepackage{fullpage}
\usepackage[T1]{fontenc}

\usepackage{epsf}
\usepackage{graphicx}
\usepackage{verbatim}
\usepackage{color}	
\usepackage{bbold}

\usepackage[most]{tcolorbox}
\usepackage{empheq}
\newtcbox{\mymath}[1][]{%
    nobeforeafter, math upper, tcbox raise base,
    enhanced, colframe=blue!30!black,
    colback=blue!30, boxrule=1pt,
    #1}

\usepackage[british]{babel}
\usepackage{verbatim}
\usepackage[T1]{fontenc}
\usepackage{lmodern}
\usepackage{lipsum}
\usepackage{booktabs}
\usepackage{caption}
\usepackage{cite}
\usepackage{soul,color}
\usepackage[toc,page]{appendix}
\usepackage{makecell}

\usepackage{ifpdf}

\usepackage{amsthm, amssymb}
\usepackage{bm}
\usepackage{mathtools}
\usepackage{amstext}
\usepackage{braket}
\usepackage{multirow}

\usepackage[normalem]{ulem}
\usepackage{xcolor}
\usepackage{cancel}

\newcommand\redsout{\bgroup\markoverwith{\textcolor{red}{\rule[0.5ex]{2pt}{0.4pt}}}\ULon}

\begin{document}
\vspace{5mm}
\vspace{0.5cm}

\def\be{\begin{eqnarray}}
\def\ee{\end{eqnarray}}

\def\ba{\begin{aligned}}
\def\ea{\end{aligned}}

\def\ls{\left[}
\def\rs{\right]}
\def\lc{\left\{}
\def\rc{\right\}}

\def\p{\partial}

\def\S{\Sigma}

\def\s{\sigma}

\def\O{\Omega}

\def\a{\alpha}
\def\b{\beta}
\def\g{\gamma}

\def\ad{{\dot \alpha}}
\def\bd{{\dot \beta}}
\def\gd{{\dot \gamma}}
\newcommand{\ft}[2]{{\textstyle\frac{#1}{#2}}}
\def\ib{{\overline \imath}}
\def\jb{{\overline \jmath}}
\def\Re{\mathop{\rm Re}\nolimits}
\def\Im{\mathop{\rm Im}\nolimits}
\def\trace{\mathop{\rm Tr}\nolimits}
\def\rmi{{ i}}

\newcommand\xrowht[2][0]{\addstackgap[.5\dimexpr#2\relax]{\vphantom{#1}}}

\def\N{\mathcal{N}}

\newcommand{\SU}{\mathop{\rm SU}}
\newcommand{\SO}{\mathop{\rm SO}}
\newcommand{\U}{\mathop{\rm {}U}}
\newcommand{\USp}{\mathop{\rm {}USp}}
\newcommand{\OSp}{\mathop{\rm {}OSp}}
\newcommand{\Symp}{\mathop{\rm {}Sp}}
\newcommand{\Sl}{\mathop{\rm {}S}\ell }
\newcommand{\Gl}{\mathop{\rm {}G}\ell }
\newcommand{\Spin}{\mathop{\rm {}Spin}}

\def\hc{c.c.}

\numberwithin{equation}{section}

\allowdisplaybreaks

\allowbreak


\hfill\text{LAPTH-017/23}

\thispagestyle{empty}
\begin{flushright}

\end{flushright}

\vspace{35pt}

\begin{center}

\def\thefootnote{\fnsymbol{footnote}}

{\LARGE \bf 
On/off scale separation 
}

\vskip 1.5cm

{\Large Fotis Farakos$^{1,2}$, \  Matteo Morittu$^{1,2}$ \ and \ George Tringas$^3$}

\vskip 1cm

$^1${\it  Dipartimento di Fisica e Astronomia ``Galileo Galilei'', Universit\`a di Padova\\ \vspace{1mm}
Via Marzolo 8, 35131 Padova, Italy}
\vspace{10pt}

$^2${\it INFN, Sezione di Padova \\ \vspace{1mm}
Via Marzolo 8, 35131 Padova, Italy}

\vspace{10pt}

$^3${\it  Laboratoire d’Annecy-le-Vieux de Physique Théorique (LAPTh); \\ 
CNRS, Université Savoie Mont Blanc (USMB), UMR 5108 \\ \vspace{1mm}
9 Chemin de Bellevue, 74940 Annecy, France}
\vspace{15pt}

\vspace{.5cm}

E-mails: fotios.farakos@pd.infn.it, matteo.morittu@pd.infn.it \& tringas@lapth.cnrs.fr

\vspace{1.3cm}

ABSTRACT 

\end{center}

We discuss minimally supersymmetric AdS$_3$ flux vacua of massive type IIA supergravity on G2-orientifolds. We find that configurations with broken scale-separation can be within finite distance from scale-separated ones, while both remain at large volume, weak coupling and have moduli stabilization. The transition is achieved with the use of a D4-brane modulus, which allows the $F_4$ flux to jump, and has an effective potential always accessible to the three-dimensional low-energy theory. Our analysis further allows us to check the distance conjecture quantitatively, as we can track explicitly the masses of the KK modes.

\thispagestyle{empty} 
\setcounter{page}{0}

\baselineskip 6mm

\newpage

\tableofcontents

\newpage




\setcounter{footnote}{0}


\section{Introduction}

Hiding the extra dimensions is one of the central requirements for achieving a realistic phenomenological scenario in string theory. 
One way to get this is to have the Kaluza--Klein modes of the internal compact space becoming very heavy compared to the external $d$-dimensional energy density 
\be
\label{SC-SEP-IN}
M_P^{d-2} \frac{m_{\rm KK}^2}{|V|} \gg 1 \,. 
\ee
This is what we typically call \textit{scale separation}. 
Clearly, having a Minkowski external space automatically implies scale separation, whereas the problem is non-trivial for (quasi) de Sitter and Anti de Sitter external spaces.

In particular, finding a scale-separated and stable de Sitter has turned out to be a very intricate problem \cite{Danielsson:2018ztv,Obied:2018sgi,Andriot:2018wzk,Garg:2018reu}, which has seen many ups and downs, starting from the original proposals that include quantum effects \cite{Kachru:2003aw,Balasubramanian:2005zx,Conlon:2005ki} and arriving until the most recent series of criticisms \cite{Sethi:2017phn,Gautason:2018gln,Gao:2020xqh,Junghans:2022exo} and alternatives \cite{Dasgupta:2019gcd,Antoniadis:2018hqy,Antoniadis:2019rkh,Brahma:2020tak,Bento:2021nbb,Bena:2022cwb}. 
Classical stable de Sitter in string theory is also a problem that can be studied independently \cite{Andriot:2020wpp,Andriot:2020vlg,Farakos:2020idt,Andriot:2022way,Andriot:2022bnb}, while scale-separated lower-dimensional de Sitter vacua are also difficult to find in extended supergravity \cite{Cribiori:2020use,DallAgata:2021nnr,Cribiori:2023ihv,Cribiori:2023gcy}. 
The actual problem becomes even harder to tackle once the unavoidable breaking of supersymmetry is also taken into account \cite{DallAgata:2022abm,Farakos:2022jcl}. 
An up-to-date review on the topic can be found in \cite{Cicoli:2023opf}. 

Understanding and establishing the existence of supersymmetric scale-separated AdS$_d$ solutions is also a question worth addressing on its own right. 
Indeed, it is important to know if a pure lower-dimensional supergravity theory can exist within a theory of quantum gravity at all, and what are its properties. 
Such scale-separated AdS vacua may also turn out to be valuable for phenomenology in the end of the day, since they can provide the basis for an uplift to de Sitter or quasi de Sitter. 
For the moment, it seems that scale-separated extended supergravity is not an option \cite{Green:2007zzb,Cribiori:2022trc,Cribiori:2023gcy}, 
whereas minimal four-dimensional supergravity is still possible \cite{DeWolfe:2005uu,Cribiori:2021djm}. 
Meanwhile, scale separation can be achieved for minimally supersymmetric AdS$_3$ \cite{Farakos:2020phe,VanHemelryck:2022ynr} from the type IIA setup, but scale separation in two dimensions is still an open problem \cite{Lust:2020npd}, and the same goes for classical scale separation in type IIB \cite{Petrini:2013ika,Emelin:2021gzx}. 
To arrive to these minimally supersymmetric models one needs orientifolds \cite{Gautason:2015tig}, whose backreaction is an open issue that can be checked with the current technology only order by order in a $g_s$ expansion \cite{Junghans:2020acz,Marchesano:2020qvg,Emelin:2022cac}. 
However, there are instances where the smearing of the sources can be shown to be reliable \cite{Blaback:2010sj,Baines:2020dmu}. 
The consistency of the minimal scale-separated AdS$_d$ vacua has been doubted as well \cite{Lust:2019zwm}, but an actual inconsistency has not been verified yet. 
Further aspects of scale-separated AdS$_d$ vacua have been studied in \cite{Conlon:2021cjk,Apers:2022zjx,Apers:2022tfm,Quirant:2022fpn,Apers:2022vfp,Plauschinn:2022ztd} with respect to the properties of their putative dual CFT; also, a discussion on the properties of the fluxes that enter such vacua can be found in \cite{Emelin:2020buq}. 
Further intricacies and difficulties in the construction of scale separation are studied in \cite{Tsimpis:2012tu,DeLuca:2021mcj,Buratti:2020kda,Font:2019uva}, whereas constructions with mild scale separation have been considered in \cite{Tsimpis:2022orc}. 
Recent developments on non-scale-separated AdS$_3$ can be found in \cite{Passias:2020ubv,Ashmore:2022kho,Couzens:2022agr,Macpherson:2023cbl}.

In the present work we will focus on the AdS$_3$ part of the story and remain also minimally supersymmetric. 
Our setup is within massive type IIA on a singular G2 orbifold and we will also include O2/O6-planes and the appropriate D-branes and fluxes for the tadpoles \cite{Farakos:2020phe}. 
Some of the O-planes will be smeared depending on the tadpole that they enter; for the moment, we can only assume that their backreaction will not be an issue. 
However, since our setup is based on \cite{Farakos:2020phe}, we do not expect a problematic backreaction at first order in $g_s$ at least \cite{Emelin:2022cac}. 
What is novel in the model that we study here is that we use a different flux/brane choice than the simple example presented in \cite{Farakos:2020phe}. 
With our choices we have the advantage to completely liberate a specific component of the $F_4$ flux from the rest ones, which are typically used to get scale separation. 
As a result, we can have a system where by changing the values of one and only component of the $F_4$ flux we can turn scale separation {\it on} and {\it off} by explicitly changing the ratio in \eqref{SC-SEP-IN}. 
In addition, what is more important is that we always remain at large volume (large radii, in particular) and weak coupling. 
Therefore, the supergravity approximation is in both cases, the scale-separated and the non-scale-separated circumstance, trustworthy. 
This is to be contrasted for example with the four-dimensional construction of \cite{DeWolfe:2005uu}, but also with the flux choice in the three-dimensional models of \cite{Farakos:2020phe}, where the non-scale-separated regime is not a good approximation for supergravity due to small volume and large $g_s$.

Once we establish the ingredients of the new flux/brane choices, we check moduli stabilization and the validity of the approximation (large volume and small $g_s$). 
Then, we further check the properties of the masses in the different regimes to see how the dimensions of their putative dual CFT operators would behave. This is an interesting aspect to test because we can compare the masses/dimensions of the very same moduli in a scale-separated and in a non-scale-separated setup. 
As a further step, we use the probe D4-brane proposed in \cite{Shiu:2022oti} to interpolate between the two vacua, and then we study their distance in moduli space. We can thus see how the masses of the KK modes change with the distance and fulfill the swampland distance conjecture \cite{Ooguri:2006in,Ooguri:2018wrx,Andriot:2020lea}.

\section{AdS$_3$ from massive IIA on G2 orientifolds} \label{Setup} 

In this section we would firstly set our conventions and briefly remind the reader the G2 scale-separated constructions of e.g. \cite{Farakos:2020phe}; then, we would introduce a new flux choice that allows, upon the variation of a flux parameter, to capture a regime where scale separation is broken.

\subsection{Basic conventions, the setup and the relevant ingredients}  

In the Einstein frame the bosonic sector of massive type IIA supergravity is described by the sum of the following NSNS and RR actions 
\begin{align}
&S_{\text{NSNS}}=\frac{1}{2\kappa^2_{10}}\int \text{d}^{10}X\sqrt{-G}\left(\mathcal{R}-\frac{1}{2}\partial_{M}\phi\partial^{M}\phi 
- \frac{1}{2}e^{-\phi}\vert H_3 \vert^2 \right) 
\, ,  \\[1mm] 
&S_{\text{RR}}=\frac{1}{2\kappa^2_{10}}\int \text{d}^{10}X\sqrt{-G}\left(-\frac{1}{2}\sum_n e^{\frac{5-n}{2}\phi}\vert F_n \vert^2 \right) \,, 
\end{align}
where the coupling $2\kappa^2_{10}$ is $2\kappa^2_{10}=(2\pi)^7\alpha^{\prime 4}$; $X^M$ denotes the ten-dimensional spacetime coordinates; the determinant of the ten-dimensional metric $G_{MN}$ is $G=\det[G_{MN}]$, and $n$ runs over $n = 0, 2, 4$. 

Then, as far as the local sources (Op-planes and Dp-branes) are concerned, 
the DBI action in the ten-dimensional language has the form
\begin{align}\label{SourceAction}
&S_{\text{Op/Dp}}=-T_p\int \text{d}^{10}X\sqrt{|G|}\,e^{-\phi}\sum_i \delta(\pi_i) \,, 
\end{align}
where we are clearly ignoring open string moduli (we will restore them only later when we work with the D4-brane). 
Within \eqref{SourceAction} the Dirac $\delta$-distribution is
\begin{align}\label{delta1}
    \delta(\pi_i)\equiv \frac{\sqrt{g_{\pi_i}}}{\sqrt{g_{7}}}\delta^{(9-p)}(y) \,,
\end{align}
where $y$ stands for the internal space coordinates; $g_{\pi_i}=\text{det}[(g_{\pi_i})_{\alpha\beta}]$ is the metric determinant of the cycle $\pi_i$ wrapped by the source; $g_{7}=\text{det}[g^{(7)}_{mn}]$ is the determinant of the internal space metric $g^{(7)}_{mn}$, and $\delta^{(9-p)}(y)$ gives the localized positions of the source in the internal space and it integrates to unit over the dual cycle $\tilde{\pi}_i$ to $\pi_i$, namely $\int_{\tilde{\pi}_i}\text{d}y^{9-p}\delta^{(9-p)}(y)=1$. Let us observe that the Dirac $\delta$-distribution depends on the internal coordinates that are transverse to the source-wrapped cycles, $y \equiv y_{\perp}$. Moreover, let us note that the tension coefficient $T_p$ is given by\footnote{The actual tension of a D-brane or O-plane contains of course both the tension coefficient $T_p$ and the $g_s^{-1}$ as in \eqref{SourceAction}.} 
\begin{align}
    T_p=N_{\rm{Op}}\mu_{\rm{Op}}+N_{\rm{Dp}}\mu_{\rm{Dp}} \quad\quad \text{with} \quad
    \mu_{\rm{Dp}}&=(2\pi)^{-p}(\alpha^{\prime})^{-(p+1)/2} \,, \quad 
    \mu_{\rm{Op}}=-2^{p-5}\times\mu_{\rm{Dp}}  \,.
\end{align}

Even though we have kept the proper powers of $\alpha^\prime$ in the expressions that we wrote so far, 
from now on, and throughout the paper we will work in string length units, i.e. 
\be
\alpha'=1 \,. 
\ee 

As far as the massive type IIA supergravity considered e.g. in \cite{Farakos:2020phe} is concerned, the relevant Bianchi identities, including the number of sources wrapping each cycle, say $N_{\rm{Op/Dp}}$, are 
\begin{align}
    \text{d}F_2& = H_3 \wedge F_0 + (2\pi)^7\left(N_{\text{O6}}\mu_{\text{O6}}\sum_i^7\delta_{i,3} 
    + \mu_{\text{D6}}\sum_i^7 N_{\text{D6}i} \delta_{i,3} \right), \label{tadpole1}\\[1mm]
    \text{d}F_6& = H_3\wedge F_4 + (2\pi)^7\left(N_{\text{O2}}\mu_{\text{O2}} 
    + N_{\text{D2}}\mu_{\text{D2}}\right)\delta_{7} \,,  \label{tadpole2}
\end{align}
for $i=1,\dots,7$. 
We have already taken into account here that the charge and tension coefficients of an O-plane are the same, that is $T_{\rm{Op}}=Q_{\rm{Op}}=\mu_{\rm{Op}}$. This clearly holds also for D-branes. Note that we will sometimes use explicitly $Q_p$ in the Bianchi identities in case we want to highlight that it is the charge that plays an important role. 
The rest of the Bianchi identities trivially vanish in \cite{Farakos:2020phe}, because there are no other sources and because $F_2=0=F_6$. 
Later on we will introduce a probe D4-brane which contributes to the Bianchi identity as 
\be\label{tadpole3}
\text{d}F_4 = H_3\wedge F_2 + (2\pi)^7  \mu_{\text{D4}} \sum_a N_{\text{D4}a} \delta_{a,5} \,,  
\ee
where the index $a$ runs over the available 2-cycles. 
This Bianchi is trivially satisfied on the SUSY-AdS$_3$ vacua that we are going to consider 
because $F_2=0$ and we do not have D4-brane sources on any vacuum. 
Indeed, the probe D4-brane ``pinches off'' when the D4-brane modulus intercepts one of the positions of the supersymmetric vacua under consideration; this will be discussed in the penultimate section of the paper.  
The quantity $\delta_{i,9-p}$ appearing in \eqref{tadpole1}, \eqref{tadpole2} and \eqref{tadpole3} is a unit-normalized $(9-p)$-form, with legs transverse to the sources wrapping $\pi_i$ and with support on the source locii. 
As we already mentioned, in the present analysis, unless otherwise noted, 
we smear the sources over the compact space in the following way: 
\begin{align}
    \delta_{i,9-p}=\delta^{(9-p)}(y)\text{d}^{9-p}y_{\perp} \quad\Rightarrow\quad
    j_{i,9-p}=\frac{\sqrt{g_{\tilde{\pi}_i}}}{\int_{\tilde{\pi}_i}\text{d}^{9-p}y_{\perp}\sqrt{g_{\tilde{\pi}_i}}}\text{d}^{9-p}y_{\perp} \,.
\end{align}

\subsubsection{The seven-dimensional internal space}

Let us now discuss the basic features of the seven-dimensional internal space $X_7$ that we would like to use for our compactification.

A general G2 structure is characterized by the invariant (under the G2 transformations) three-form 
\be
\label{Phi-rad}
\Phi = e^{127}
-e^{347}
-e^{567} 
+e^{136}
-e^{235}
+e^{145}
+e^{246} \,,
\ee
where $e^{ijk}=e^{i}\wedge e^{j}\wedge e^{k}$ and $e^i$ is the co-vielbein that makes up the internal metric. 
Flux compactifications on G2 spaces have been further discussed for example in \cite{Beasley:2002db,DallAgata:2005zlf,Derendinger:2014wwa,Danielsson:2014ria} (spaces with G2-holonomy where constructed in \cite{Joyce}). 
We can then introduce the orthonormal basis $\text{d}y^m$ for the seven-dimensional space $X_7$ 
and define an invariant basis (under the orbifold action that we are going to specify in a while) of harmonic three-forms
\begin{align}\label{3formbasis}
    \Phi_i = \{+\text{d}y^{127}, -\text{d}y^{347}, -\text{d}y^{567}, +\text{d}y^{136}, -\text{d}y^{235}, +\text{d}y^{145}, +\text{d}y^{246}\} \,, 
\end{align}
for $i=1,\dots,7$. The Hodge dual of the fundamental 3-form $\Phi$ is
\begin{align}\label{4formbasis}
\star\Phi = e^{3456}
-e^{1256}
-e^{1234} 
+e^{2457}
-e^{1467}
+e^{2367}
+e^{1357} \,,
\end{align}
which can be expanded (similarly to the 3-form \eqref{Phi-rad}) in the basis of invariant 4-forms
\begin{align}
    \Psi_i = \{+\text{d}y^{3456}, -\text{d}y^{1256}, -\text{d}y^{1234}, +\text{d}y^{2457}, -\text{d}y^{1467}, +\text{d}y^{2367}, +\text{d}y^{1357}\} \, , 
\end{align} 
for $i=1,\dots,7$. 
The invariant 3-form and 4-form basis elements $\Phi_i$ and $\Psi_i$ satisfy $\int \Phi_i\wedge \Psi_j = \delta_{ij}$. 
The spaces with G2-holonomy we will focus on are characterized by vanishing exterior derivatives of the fundamental forms, i.e.
\begin{align}
    \text{d}\Phi=0 \,, \quad\quad \text{d}\Psi=0 \,.
\end{align}
More precisely, 
we choose our G2-holonomy internal space to be a seven torus $T^7$ with the following periodical identification: 
\begin{align}
    y^m \simeq y^m+1 \quad \text{for } m=1,\dots,7 \,.
\end{align}
The vielbeins in the fundamental forms are then directly related to the radii of the torus, namely $e^m=r^m \text{d}y^m$, where $r^m$ stands for each of the radii of the corresponding seven cycles. The fundamental 3-form can thus be written as
\begin{align}
    \Phi = \sum_is^i\Phi_i \,, 
\end{align}
where the $s^i$'s are some moduli that describe deformations of the internal space. Once expressed in terms of the radii, these moduli have, for instance, the form $s^1 = r^1 r^2 r^7$.
The seven-dimensional torus $T^7$ with G2-holonomy just described is invariant under specific $Z_2$ involutions, which are given by the group of isometries $\Gamma = \{\Theta_{\alpha},\Theta_{\beta},\Theta_{\gamma}\}$ and lead to the toroidal orbifold  
\begin{equation}\label{orbifold}
    X_7 = \frac{T^7}{Z_2 \times Z_2 \times Z_2} \,. 
\end{equation} 
The action of these three involutions on the internal coordinates is 
\be
\begin{aligned}\label{Z2s}
\Theta_\alpha : y^i & \to (-y^1, -y^2, -y^3, -y^4, y^5, y^6, y^7) \,, 
\\[0.5mm]
\Theta_\beta : y^i & \to (-y^1, -y^2, y^3, y^4, -y^5, -y^6, y^7) \,,
\\[0.5mm]
\Theta_\gamma : y^i & \to (-y^1, y^2, -y^3, y^4, -y^5, y^6, -y^7) \,, 
\end{aligned}
\ee
while the combined involutions of the orbifold group $\Gamma$ also preserve both the calibration $\Phi$ and its Hodge dual $\star \Phi$: they are
\be
\begin{aligned}
\Theta_{\alpha}\Theta_{\beta} : y^i & \to (y^1, y^2, -y^3, -y^4, -y^5, -y^6, y^7) \,, 
\\[0.5mm]
\Theta_{\beta}\Theta_{\gamma} : y^i & \to (-y^1, -y^2, y^3, y^4, -y^5, -y^6, y^7) \,,
\\[0.5mm]
\Theta_{\gamma}\Theta_{\alpha} : y^i & \to (y^1, -y^2, y^3, -y^4, -y^5, y^6, -y^7) \,, 
\\[0.5mm]
\Theta_{\alpha}\Theta_{\beta}\Theta_{\gamma} : y^i & \to (-y^1, y^2, y^3, -y^4, y^5, -y^6, -y^7) \,. 
\end{aligned}
\ee 
Since the presence of O2-planes is also needed for the cancellation of the relevant tadpole, we are not actually done with the involutions yet. The action on the internal coordinates of the $Z_2$ involution $\sigma$ associated with O2-planes is specified by 
\be
\sigma : y^i\rightarrow -y^i \,.
\ee
The involutions defining the isometry group $\Gamma$ combine with $\sigma$ and give three images of the form $\Theta_{\alpha}\sigma$, three other images of the form $\Theta_{\alpha}\Theta_{\beta}\sigma$ and another one that is $\Theta_{\alpha}\Theta_{\beta}\Theta_{\gamma}\sigma$. The positions that remain invariant under the involutions considered just above are interpreted as the localized positions of O6-planes. The presence of both the involutions related to $\Gamma$ and $\sigma$ is essential for a proper reduction of supersymmetry: it allows to truncate the thirty-two potential supercharges down to two real supercharges, leading to a three-dimensional effective theory with N=1 supersymmetry. For details on the AdS$_3$ N=1 supergravity setup see e.g. \cite{Farakos:2020phe}.

The volume of the internal space is defined as follows 
\begin{align}\label{volume}
    \text{vol}(X_7)=\left(\prod_{i=1}^7s^i\right)^{1/3}=\frac{1}{7}\int \Phi\wedge \star\Phi \,, 
\end{align}
where we normalize $\int_7 \text{d}y^{1234567}=1$. The second equality in \eqref{volume} is general for spaces which admit G2-structure. We would also like to recall the useful Hodge dual expressions
\be
\star \Phi = \sum_i \frac{\text{vol}(X_7)}{s^i} \Psi_i \,, \quad\quad \star \Phi_i = \frac{\text{vol}(X_7)}{(s^i)^2} \Psi_i \,. 
\ee
As far as the internal space is concerned, we simply consider a seven-dimensional torus whose line-element is given by
\begin{align}\label{lineelement}
    \text{d}s^2_7=\sum_{i=1}^7 r_i^2 \text{d}y_i^2 = \rho \sum_{i=1}^7\tilde{r}_i^2 \text{d}y_i^2 \equiv \rho \text{d}\tilde{s}_7^2 \,,
\end{align} 
where $\text{d}\tilde{s}_7^2$ is the metric on the corresponding unit-volume G2-holonomy space and $\rho$ the volume modulus. After the volume dependence has been extracted, the deformation moduli $s^i$ take the form
\begin{align}
    s^i=\text{vol}(X_7)^{3/7}\tilde{s}^i \, ,
\end{align}
and, once the unit volume condition in \eqref{volume} is imposed, the following relation for the moduli is obtained:
\begin{align}\label{unitvolume}
\text{vol}(\tilde{X}_7)=1 \quad \Rightarrow \quad \prod_{i=1}^7 \tilde s_i = 1 \,.
\end{align}
This relation should be accounted for when counting the independent degrees of freedom at unit volume.

\subsubsection{The three-dimensional effective action}

The metric ansatz that we use for the reduction to three dimensions is 
\be\label{metricansatz}
\text{d}s_{10}^2 = e^{2 \alpha v} \text{d}s_3^2 + e^{2 \beta v} \text{d}\tilde s_7^2 \,. 
\ee
For $\alpha^2=7/16$ and $-7\beta=\alpha$ we get canonical kinetic terms for the universal moduli in the three-dimensional theory, namely
\begin{align}
   (2 \pi)^7 e^{-1} \mathcal{L}=  R_3 - \frac{1}{2}(\partial \upsilon)^2-\frac{1}{2}(\partial \phi)^2-V_{\rm dim.} \, . 
\end{align}
Here $e$ indicates the determinant of the three-dimensional external space, $e = \sqrt{-\text{det}[g^{(3)}]}$, 
which we hope that will not confuse the reader since the same symbol has been also used for other purposes. 
For this metric ansatz the individual internal space radii are $r_i = e^{\beta v} \tilde r_i$ such that
\begin{align}
    s^i=e^{3\beta\upsilon}\tilde{s}^i \quad\quad \text{and} \quad\quad \text{vol}(X_7)=e^{7\beta \upsilon} \,.
\end{align}
In order to match with the conventions for three-dimensional supergravity theories in the literature, we perform a rescaling to the three-dimensional metric $g_{\mu\nu} \rightarrow \frac{(2 \pi)^{14}}{4}g_{\mu\nu}$. The three-dimensional supergravity Lagrangian is
\be
e^{-1} {\cal L} = \frac12 R_3 - G_{IJ} \partial \varphi^I \partial \varphi^J - V \,, 
\ee
with the scalar potential given by
\begin{align}\label{scalarpotentialsugra}
    V=G^{IJ}P_IP_J-4P^2 \,,
\end{align}
with $P_I=\partial_I P$ representing the derivative of the superpotential $P$ with respect to the moduli $\varphi^I$, 
and $G^{IJ}$ being the inverse scalar field metric (see e.g. \cite{Farakos:2020phe}). 
The potential obtained from dimensional reduction is related to \eqref{scalarpotentialsugra} via $V_{\rm dim.} = 8V$. 

We decide to perform the linear transformation onto the universal moduli
\begin{align}
    7\beta\upsilon = \frac{\sqrt{7}}{32}x-\frac{21}{32}y \,, \quad\quad \phi=-\frac{3\sqrt{7}}{8}x-\frac{1}{8}y \,,
\end{align}
so that the kinetic terms of the final three-dimensional effective action are given in \cite{Farakos:2020phe} as follows
\begin{align}
    e^{-1}\mathcal{L}_{\text{kin.}}=
    \frac{1}{2} {R}_3
    -\frac{1}{4} (\partial x)^2
    -\frac{1}{4} (\partial y)^2
    -\frac{1}{4}\text{vol}(\tilde{X}_7)^{-1}\int_7\Phi_i\wedge \tilde{\star}\Phi_j \partial_{\mu} \tilde{s}^i \partial^{\mu} \tilde{s}^j  \,.
\end{align}
The K\"ahler metric $G_{IJ}$ can be read off from this expression. In particular, evaluating the integral over the internal space and expressing one of the deformation moduli, e.g. $\tilde{s}^7$, in terms of the remaining six independent fields $\tilde{s}^{a}$, the kinetic terms can be rewritten as
\begin{align}
    \frac{\delta_{ij}}{4 \tilde{s}^i \tilde{s}^j} \partial_{\mu} \tilde{s}^i \partial^{\mu} \tilde{s}^j \,\equiv\, \tilde{G}_{ab}\, \partial_{\mu}\tilde{s}^a\partial^{\mu}\tilde{s}^b \quad\Rightarrow\quad \tilde{G}_{ab} = \frac{1+\delta_{ab}}{4\tilde{s}^a\tilde{s}^b} \,,
\end{align}
for $a,b=1,\dots,6$, and the field space metric for the $\tilde{s}^a$ moduli consequently shows up (see e.g. \cite{Emelin:2021gzx}). The scalar field Lagrangian contains eight \textit{independent} moduli (the universal $x$ and $y$ and the six deformations $\tilde{s}^a$) and its final form becomes
\begin{align}
    e^{-1}\mathcal{L}=
    \frac{1}{2} {R}_3
    -\frac{1}{4} (\partial x)^2
    -\frac{1}{4} (\partial y)^2
    -\frac{1+\delta_{ab}}{4\tilde{s}^a\tilde{s}^b} \partial_{\mu} \tilde{s}^a \partial^{\mu} \tilde{s}^b  -V(x,y,\tilde{s}^a) \,.
\end{align}

\subsection{Scale separation and radii} \label{SSandR}

In this subsection we would like to discuss how exactly scale separation works. In particular, we present a complementary derivation of the scale separation condition used in \cite{Farakos:2020phe}, which allows us to deduce the condition for the breaking of scale separation by individual radii.

To this purpose we only need to focus on the dilaton scalar KK modes and see their influence from the metric radii. Because of our toroidal ansatz for the metric of the internal seven-dimensional space the period of the internal dimensions can be extracted from the identifications $y^m \simeq y^m + 1$. We can then study the masses of the KK modes related to the dilaton $\phi$. More precisely, we can concentrate on the KK modes related to a specific compact internal dimension, say $r_1 \equiv R = e^{\beta v} \tilde R$; we furthermore refer to the internal coordinate of interest as $y_1 \equiv Y$. In other words, our metric ansatz in \eqref{metricansatz} takes the form 
\be
\text{d}s_{10}^2 = e^{2 \alpha v} \text{d}s_3^2 + e^{2 \beta v} \tilde R^2 \text{d}Y^2 + \dots \,.
\ee
The periodicity of the internal space coordinate being accounted for, upon dimensional reduction we have that $\phi(x, Y + 1) = \phi(x,Y)$; the KK modes of the three-dimensional theory related to the specific chosen dimension are then
\be
\phi(x,Y) = \sum_{n} \phi_n(x) \cos[2 \pi n Y] \,. 
\ee
We now want to look at the masses of the three-dimensional $\phi_n$ fields and compare them to the vacuum energy. In particular, we can simply look at the mass of the scalar $\phi_1(x)$, which we call $u(x)=\phi_1(x)$, and we consequently work with the ansatz 
\be
\phi(x,Y) = \phi(x) + u(x) \cos[2 \pi Y] \,. 
\ee
Here we abuse notation and use the same letter for the three-dimensional dilaton $\phi(x)$ and for the original ten-dimensional one $\phi(x,Y)$; we hope that this does not confuse the reader as it will be clear which one we are referring to depending on the context. 
Then, we proceed with the dimensional reduction\footnote{Let us keep in mind that 
$\int_0^1 \cos[2 \pi Y] \text{d}Y = 0$\,, 
$\int_0^1 (\cos[2 \pi Y])^2 \text{d}Y = 1/2$\,, 
$\int_0^1 \cos[2 \pi Y] \sin[2 \pi Y] \text{d}Y=0$ and 
$\int_0^1 (\sin[2 \pi Y])^2 \text{d}Y=1/2$\,.}
to find 
\be
\begin{aligned}
S_{\text{NSNS}} =& \,\frac{1}{(2 \pi)^7} \int \text{d}^3x \, \text{d}Y \sqrt{-g_3} \, e^{3 \alpha v + 7 \beta v} 
\Big{[} e^{-2 \alpha v}  R_3 
- \frac12  e^{-2 \alpha v} g^{\mu \nu} \partial_\mu \phi \partial_\nu \phi +
\\[1mm]&
- \frac12  e^{-2 \alpha v} (\cos[2 \pi Y])^2 g^{\mu \nu} \partial_\mu u \partial_\nu u 
- \frac12  e^{-2 \beta v} \frac{1}{\tilde R^2} (2 \pi \sin[2 \pi Y])^2 u^2  
\Big{]} + \dots \,,
\end{aligned}
\ee
from which we deduce 
\be
m_u^2 = (2 \pi)^2 \frac{e^{2 \alpha v - 2 \beta v}}{\tilde R^2} \,. 
\ee
The comparison of the mass of the given KK mode to the vacuum energy is made by simply taking the ratio $\langle V \rangle /m_u^2$, 
and asking, in order to have scale separation, that this ratio goes to zero. This means, in fact, that $m_u^2 \gg |\langle V \rangle| \sim L_{\Lambda}^{-2}$, the Planck mass being a constant. $L_\Lambda$ stands for the AdS$_3$ radius. Specifically, because $\alpha = - 7 \beta$, we evaluate 
\be
\frac{\langle V \rangle}{m_u^2} = \frac{\tilde R^2}{(2 \pi)^2} \times \langle V \rangle \times e^{2 \beta v - 2 \alpha v} 
\sim \tilde R^2 e^{16 \beta v} \langle V \rangle \,. 
\ee
We see that, when $\tilde R$ does not scale, then the criterion for scale separation is given by
\begin{align}
    e^{16 \beta v} \langle V \rangle \to 0  \,.  
\end{align}
If the $\tilde R$ scales, instead, then it should be included in the scaling that defines the scale separation regime.

\subsection{Fluxes, tadpoles and scalings} 

After having discussed how scale separation is realized, we would like to introduce a new flux choice that will be the main topic of the upcoming discussion. 
Our ansatz here is different from the isotropic ansatz studied in \cite{Farakos:2020phe}. 
We will insert a stack of D6-branes on the cycle $\Phi_7$ therefore the $H_3$ flux along the $\Phi_i$ basis is 
\be\label{HAnsatz}
H_3 = \sum_{i=1}^7 h^i \Phi_i \,, 
\ee
with
\begin{align}\label{Hflux}
    h^i = h (1,1,1,1,1,1,0) \,. 
\end{align}
The $F_4$ flux expanded on the $\Psi_i$ basis is 
\be
F_4 = \sum_{i=1}^7 f^i\,\Psi_i  \,, 
\ee 
which can be written as
\be\label{F4Expansion}
F_4 = F_{4,q} + F_{4,f} = \sum_{i=1}^7 f^i_{4,q}\,\Psi_i + \sum_{i=1}^7 f^i_{4,f}\,\Psi_i  \,
\ee
with
\be\label{F4Ansatz}
f_{4,q}^i = q (0,0,0,0,0,0,-1) \,, \quad\quad f_{4,f}^i = f (-1,-1,-1,-1,-1,5,0) \,.
\ee
This flux choice admits a scale-separated solution. Moreover, when the $F_4$ flux is changed, 
for instance by moving $q$ consistent with flux quantization as 
\be
q \to q - (2 \pi)^3 \,. 
\ee
We are going to show that starting from a non-scale-separated solution and performing such many steps, such that $q \sim (2 \pi)^3 \times 1$, 
we arrive at a scale-separated regime and the refined distance conjecture can be explicitly realized. 

Let us see how the parameters $q$ and $f$ remain unbounded and the tadpole is cancelled. The integral of Bianchi identities over the compact space should vanish. In our case, the tadpole of $F_6$ cancels by choosing appropriate numbers of fluxes and sources within
\begin{equation}
    \int_7\text{d}F_{6} = 0 = \int_7 H_3\wedge F_4 + \int_7(2\pi)^7(N_{\rm{O2}}\mu_{\rm{O2}}+N_{\rm{D2}}\mu_{\rm{D2}})\delta_{7} \,.
\end{equation}
The number of Op-planes depends on the number of fixed points that the orientifold involution has in the internal manifold. For the toroidal orbifold \eqref{orbifold} the number of O2-planes is related to the $N_{\text{O2}}=2^7$ corresponding orientifold fixed points, while for a single O6-plane $N_{\text{O6}}=2^3$ for each 3-cycle. 
For such a cancellation we have 
\begin{align}
-2^{-3}N_{\rm{O2}}+N_{\rm{D2}}=0 \,,
\end{align}
which holds for $N_{\rm{D2}}=2^4$, and, integrating over the internal space,
\begin{align}
    \int_7H_3\wedge F_{4,q}&=0\times (-q)=0 \,, \\[1mm]
    \int_7H_3\wedge F_{4,f}&=-5hf+5hf=0 \,.
\end{align}
The flux quanta $q$, $h$ and $f$ can always be chosen to be integers up to appropriate $2\pi$ factors (see Appendix \ref{FluxAppendix} for conventions on flux quantization). We cancel the tadpole of $F_2$ (which is identically zero in a setup where $F_2=0$) by integrating over 3-cycles  
\begin{equation}
    \int_3\text{d}F_{2} = 0 = \int_3 H_3\wedge F_0 + \int_3(2\pi)^7(N_{\rm{O6}}\mu_{\rm{O6}}+N_{\rm{D6}}\mu_{\rm{D6}})\delta_{3} \,.
\end{equation}
Because of our flux ansatz \eqref{HAnsatz} we need to cancel the orientifold with a proper flux choice and place a specific number of D6-branes on the cycle $\Phi_7$ to cancel the relevant orientifold. We integrate over each cycle $\Phi_i$ for $i=1,\dots,6$, to get 
\begin{align}\label{tad2}
    h m - 2\pi N_{\rm{O6}} = 0 \,.
\end{align}
For the local objects wrapping the cycle $\Phi_7$ and after integration 
\begin{align}
- 2N_{\rm{O6}}+N_{\rm{D6}}=0 
\end{align}
holds, which implies $N_{\rm{D6}}=2^4$, since we have $N_{\rm{O6}}=2^3$ per 3-cycle.

\subsubsection{The scaling of the fields}

The general superpotential of the three-dimensional theory that we obtain upon compactification on $X_7$ is \cite{Farakos:2020phe}
\be 
\label{superpotential}
\frac{P}{(2\pi)^7} = \frac{e^y}{8} \left[e^{\tfrac{x}{\sqrt{7}}} \int \star\Phi \wedge H_3 \, \text{vol}(X_7)^{- \frac47} 
+ e^{-\tfrac{x}{\sqrt{7}}} \int \Phi \wedge F_4 \, \text{vol}(X_7)^{- \frac37}  \right] +  \frac{F_0}{8} \, e^{\tfrac12 y - \tfrac{\sqrt{7}}{2} x } \, ,
\ee
in $M_P$ units. 
Once we insert the flux choices \eqref{HAnsatz} and \eqref{F4Ansatz} the superpotential becomes
\be\label{SSRsuperpotential}
\frac{P}{(2\pi)^7} = \frac{e^{y+\frac{x}{\sqrt{7}}}}{8} h \sum_{i=1}^6 \frac{1}{\tilde{s}^i} + \frac{e^{y-\frac{x}{\sqrt{7}}}}{8} \left[ 
f \left(-\sum_{i=1}^5 \tilde{s}^i + 5\tilde{s}^6 \right) - q \tilde{s}^7 \right] + \frac{m}{8} e^{\frac{y-\sqrt{7}x}{2}}\,,
\ee
while from the unit volume in \eqref{unitvolume} we have
\begin{align}
    \tilde{s}^7 = \prod_{a=1}^6 \frac{1}{\tilde{s}^a} \,,
\end{align}
with $a=1,\dots,6$. 
The scaling for the fields is 
\begin{align}
    f \sim N \,,\quad 
    e^y \sim N^Y \,,\quad 
    e^x \sim N^X \,, \quad 
    \tilde{s}^{\alpha} \sim N^S \,, \quad 
    q  \sim  N^Q \,,
\end{align}
while the scaling of $h$ and $m$ is fixed by the O6 Bianchi identity to be $h \sim N^0$ and $m \sim N^0$ since the smeared sources do not scale $j_{i,3}\sim N^0$. 
In other words we see how the other fields and fluxes behave when we increase $f$ (always respecting flux quantization for $f$ and $q$ of course). 
Asking that all terms in $P$ scale in the same way we find that 
\begin{align}
    Y = -\frac{9}{2} - 7S \,,\quad\quad 
    X = \frac{\sqrt 7}{2}(1 + 2S) \,,\quad\quad 
    Q = 1 + 7S \,,  
\end{align}
and $P \sim N^{-4 - 7S}$.
From the scalings of the fields we also find 
\begin{align}\label{vscaling}
  e^{7 \beta v} \sim  \text{vol}(X_7) \sim N^{\frac{\sqrt{7}}{32}X-\frac{21}{32}Y}=N^{\frac{7}{16}(7+11S)} \quad \Rightarrow \quad e^{\beta v} \sim N^{\frac{7+11S}{16}} \,.
\end{align}
It is worth to observe that for $S=0$ a model behaving similarly to the isotropic model of \cite{Farakos:2020phe} is reproduced: the internal space volume, for instance, scales like $\text{vol}(X_7)\sim N^{49/16}$.  

From the internal space seven-dimensional metric \eqref{lineelement} we can introduce the length scales
\be
    L_{\text{KK},i}^2 = \rho\tilde{r}_i^2 \,.
\ee
Following the considerations of subsection \ref{SSandR}, we would like to extract the scaling of the ratios $L_{\text{KK},i}^2/L_{\Lambda}^2$. To this purpose, we should first discuss the scaling of the moduli fields $\tilde{r}_i$. Let us recall that 
\begin{align} \label{stildescaling}
 \tilde{s}^{a} &\rightarrow \tilde{s}^{a} \times N^S   \quad \text{(for } a = 1,\dots,6 \text{)} \,, \\[0.5mm]
 \tilde{s}^7 &\rightarrow \tilde{s}^7 \times N^{-6 S} \,.
\end{align}
Then, the radii $\tilde{r}_i$ are such that
\begin{align}
&\tilde{r}^1 \sim \tilde{r}^3 \sim \tilde{r}^5 \sim \tilde{r}^7 \sim  N^{3S/2} \,,  \\[0.5mm]
&\tilde{r}^2 \sim \tilde{r}^4 \sim \tilde{r}^6 \sim  N^{-2 S} \,, 
\end{align}
and the radii $r_i = \rho\,\tilde{r}_i$ scale as
\begin{align} \label{r246scaling}
r_{1,3,5,7}^2 &= e^{2 \beta v} \tilde r_{1,3,5,7}^2 \sim N^{\frac{7 + 11 S}{8}} \times N^{3 S} \,,  \\[0.5mm]
r_{2,4,6}^2 &= e^{2 \beta v} \tilde r_{2,4,6}^2 \sim N^{\frac{7 + 11 S}{8}} \times N^{-4 S} \,. 
\end{align}
We would like to set our discussion in a large volume regime, which can be achieved when 
\be
r_i^2 = \rho \tilde r_i^2 = e^{2 \beta v} \tilde r_i^2 \gg 1 \quad \text{for any } i \,, 
\ee
and, as a consequence 
\be \label{LargeVolume}
\frac{7 + 11 S}{8} + 3 S > 0 \quad \& \quad 
\frac{7 + 11 S}{8} -4 S > 0 \quad \Rightarrow \quad 
- \frac15 < S < \frac13 \,. 
\ee
Other than the large volume condition\footnote{Note that we are requiring a refined large volume condition here, 
that is, not only the total volume of the internal space is large, 
but each one radius is also large compared to the string length.}, 
we would also like the string coupling to be small. 
Therefore, being
\be \label{gsscaling}
g_s = e^\phi \sim N^{-\frac{3 + 7S}{4}} \,, 
\ee
we should require that 
\be \label{WeakCoupling}
\frac{3}{4} + \frac74 S > 0 \quad \Rightarrow \quad S > - \frac37 \,. 
\ee
As the reader can appreciate, once the large volume limit is respected, the string coupling is ensured to be small. 

This being discussed, we could now check if scale separation can be realized or broken in the model. 
That is we want to check if first of all the radii can be parametrically small compared to the AdS radius, 
but still remain at large volume and weak coupling. 
Then we will check the breaking of scale separation, 
that is, 
to check if any of the radii can become larger or comparable to the AdS length scale while the volume 
is still large and the string coupling is small.

The internal space radii split in two main categories, which we will consider separately. 
To estimate scale separation we first note that the AdS$_3$ radius $L_\Lambda$ scales as 
\be
\frac{1}{L_{\Lambda}^2} \sim P^2 \sim N^{-8 - 14S} \, , 
\ee
because in 3D supergravity $V = -4 P^2$. 
The radii $\{r_i\}_{i=1,3,5,7}$ are such that 
\be
    \frac{L_{\text{KK},i}^2}{L_{\Lambda
    }^2} \sim e^{16\beta\upsilon} V \times \tilde{r}^2_1 
    \sim N^{7 + 11S} \times N^{-8 -14S} \times N^{3S} = N^{-1} \quad \text{for }  i=1,3,5,7 \,\,,
\ee
interestingly these ratios do not develop any dependence on $S$ and can not be responsible for the potential breaking of scale separation. 
Therefore these radii always realize parametric scale separation for large $N$. 
The remaining radii $\{r_i\}_{i=2,4,6}$, instead, give 
\be \label{SSBr246}
    \frac{L_{\text{KK},i}^2}{L_{\Lambda
    }^2} \sim e^{16\beta\upsilon} V \times \tilde{r}^i_2 
    \sim N^{7 + 11S} \times N^{-8 -14S} \times N^{-4S} = N^{-1 - 7 S} \quad \text{for }  i=2,4,6 \, 
\ee
and parametric scale separation is realized for $S>-1/7$ for large $N$, 
whereas it can be broken if 
\be
S \leq - \frac{1}{7} \,, 
\ee
which has a non-vanishing intersection with the conditions in \eqref{LargeVolume} and \eqref{WeakCoupling}. 
In other words, 
we have just shown that the breaking of scale separation in a large volume regime at weak string coupling is achievable as long as 
\be
- \frac{1}{5} < S \leq - \frac{1}{7} \,. 
\ee
Instead, for example when $S=0$ we have scale separation.

In the present work we present an interesting example, for which we are going to take
\be \label{SSSB}
S = - \frac17 (1 + \epsilon) \,, \quad\quad  1 \gg \epsilon > 0 \,, 
\ee
i.e. $S < - \frac17$ but infinitesimally near the limiting value $-\frac17$, and there study the breaking of scale separation and the realization of the distance criteria. 
Even though the reader should keep in mind that $S < - \frac17$, for practical reasons we are going to refer the following discussion to the limiting case 
\be
S = - \frac17  \,, 
\ee 
and only when $\epsilon$ is essential in the discussion we will bring it up. 
When $S = -\frac17$, \eqref{SSBr246} implies that $\{L_{\text{KK},i}^2/L_{\Lambda}^2\}_{i=2,4,6}$ do not scale with $N$, or equivalently that $\{L_{\text{KK},i}^2\}_{i=2,4,6}$ and $L_{\Lambda}^2$ have become comparable in magnitude, thus violating scale separation. Moreover, as $S = - \frac{1}{7}$,
\be
Q = 0 \,, 
\ee
which suggests that we can travel between the broken scale-separated to scale-separated regime by changing $Q$ from $Q=0$ to $Q=1$.

Specifically, we can imagine to start from a background with a small amount of $F_4$ flux, where $f$ and $q$ scale differently and scale separation is broken but the system remains in a large volume regime at weak string coupling. Then, as we will see soon, via a D4-brane (or a stack of D4-branes), we can increase the initially large value of $q$, which is quantized, up to $q \sim (2 \pi)^3 \times 1$. Once scale separation is restored 
as $q \sim f$, the system remains in a large volume regime at weak string coupling. 
This is crucial in order to be able to still trust the compactification. 
This property also makes the present construction different with respect to the DGKT model \cite{DeWolfe:2005uu}. 
It is also important to stress that the potential of the open string modulus 
that allows to interpolate between the broken scale-separated regime and the circumstance where scale separation is realized 
can be always be made parametrically smaller than the flux potential (we will verify this in the appropriate section). 
Moreover, when scale separation is completely broken, towers of KK modes associated with the radii $\{r_i\}_{i=2,4,6}$ become light. This amounts to an explicit realization of the refined distance conjecture, as these towers appear when a finite distance is travelled in field space by the modulus representing the position of the D4-branes. 
In particular, as we will also explain in Section \ref{D4}, we can explicitly check that 
\be \label{DCformula}
m \sim m_0 e^{- \gamma \Delta } \,, 
\ee
where $\Delta$ is the above mentioned distance and $\gamma$ is ${\cal O}(1)$ coefficient.

\section{Moduli stabilization}

\subsection{The two regimes}

In this section we would like to show that the closed string moduli are stabilized for our flux choices. 
In particular, by solving the supersymmetry conditions 
\be
\frac{\partial P}{\partial x} =  \frac{\partial P}{\partial y} = \frac{\partial P}{\partial \tilde s^a} =0 
\ , \quad a=1,\dots,6\,, 
\ee
we are going to show that such stabilization can be achieved both for the circumstance where scale separation is realized and the case for which it is broken. 
Because $P$ gets a non-trivial value on the vacuum we have SUSY-AdS$_3$.

\subsubsection{The scale-separated regime}\label{scaleseparated}

Let us first consider the regime where scale separation is realized, i.e. $S=0$ or $Q=1$. 
In the presence of scale separation we can choose $q$ to be $q=f$ in \eqref{SSRsuperpotential}. 
We would like to search for solutions of the form
\be \label{SSRAnsatz}
\langle \tilde{s}^a \rangle_{a=1,...,5} = \sigma \,, 
\quad\quad \langle \tilde{s}^6 \rangle = \tau 
\quad\quad \text{and} 
\quad\quad \langle \tilde{s}^7 \rangle = \frac{1}{\sigma^5 \tau} \,.
\ee
Once taking the derivatives of the superpotential \eqref{SSRsuperpotential} when $q=f$ and invoking the ansatz \eqref{SSRAnsatz}, the equations we are led to are
\be
1 - a \sigma^5 + 5 \sigma^5 \tau^2 = 0 \,, \quad\quad
1 - a \sigma^4 \tau - \sigma^6 \tau = 0 \,,
\ee
and
\be
- 3 b + 2 a \left( \frac{5}{\sigma} + \frac{1}{\tau} \right) = 0 \, ,
\quad\quad
\frac{b}{2} + a \left(\frac{5}{\sigma} + \frac{1}{\tau}\right) + \left(-5\sigma +5\tau - \frac{1}{\sigma^5\tau}\right) = 0 \,,
\ee
where we have defined
\be
a =\frac{h}{f} e^{\frac{2x}{\sqrt{7}}}\,, \quad\quad
b =\frac{m}{f}  e^{-\frac{y}{2}-\frac{x}{2\sqrt{7}}}\,.
\ee
We can solve the previous equations either analytically or numerically. The numerical evaluation gives
\be
a = 0.576975\dots\,, 
\quad b = 3.39698\dots\,,
\quad \sigma = 1.22904\dots\,, 
\quad \tau = 0.209945\dots\,.
\ee
We thus show that in the scale-separated regime the closed string moduli are stabilized.

\subsubsection{The non-scale-separated regime}\label{nonscaleseparated}

Let us now repeat the calculation that we have just described in the regime where scale separation is broken, i.e. when $S=-\frac{1}{7}$ or $Q=0$. 
Once we define the parameter $c$ that controls the ratio of the two types of $F_4$ flux as 
\be
c = \frac{q}{f} \,, 
\ee
the extremization equations become
\be
c - a \sigma^5 + 5 \sigma^5 \tau^2 = 0 \,, \quad\quad
c - a \sigma^4 \tau - \sigma^6 \tau = 0  \,, 
\ee 
and
\be
- 3b + 2a \left( \frac{5}{\sigma} + \frac{1}{\tau} \right)=0 \,,
\quad\quad \frac{b}{2} + a \left( \frac{5}{\sigma} + \frac{1}{\tau} \right) + \left( -5 \sigma + 5 \tau - \frac{c}{\sigma^5 \tau} \right)=0 \,. 
\ee 
We can solve these equations numerically for different values of $c$, e.g. $c = 10^{-1}, \, 10^{-3}, \, 10^{-6}, \, 10^{-9} $, and show that, as long as $q \ne 0$, moduli stabilization can again be achieved. 
Small values for the parameter $c$ can be obtained for large \textit{integer} (always times $(2 \pi)^3$) 
values of $f$ such that the flux parameters $q$ and $f$ appearing in the effective theory are always 
integers ($\times (2 \pi)^3$) thus properly quantized. 
The results are summarized in Table \ref{CCtable}.
\begin{center}\label{table1}
\renewcommand{\arraystretch}{1.4}
\begin{tabular}{|c||c|c|c|c|}
     \hline 
     $c$ & $a$ & $b$ & $\sigma$ & $\tau$  \\ 
     \hline \hline
     $10^{-1}$ & $0.298843$  &  $2.44476$ &  $0.884523$ &  $0.151095$  \\
     \hline
     $10^{-3}$ & $0.0801704$  &  $1.26626$ &  $0.458136$ &  $0.078259$ \\
     \hline
     $10^{-6}$ & $0.0111396$  &  $0.472009$ &  $0.170775$ &  $0.0291718$ \\
     \hline
     $10^{-9}$ & $0.00154785$  & $0.175946$ & $0.0636578$ &  $0.0108741$ \\
     \hline
\end{tabular}
\captionof{table}{\label{CCtable} The table shows the values of $a$, $b$, $\sigma$ and $\tau$ for which moduli stabilization can be achieved as the parameter $c$ takes the values $10^{-1}$, $10^{-3}$, $10^{-6}$ and $10^{-9}$.}
\end{center}

\subsection{Masses and dimensions of dual operators} 

After having discussed the stabilization of the closed string moduli both in the circumstance where scale separation is realized ($c = \frac{q}{f} = 1$) and in the case where it is broken ($c = \frac{q}{f} \ll 1$), we can turn to the evaluation of their squared scalar masses. In the scale-separated regime all the masses are positive and the same happens when scale separation is broken. 

The scalar potential is found from \eqref{scalarpotentialsugra} to be
\begin{equation}\label{potential}
V =\, F(\tilde{s}^a)e^{2y-\frac{2}{\sqrt{7}}x}  
+H(\tilde{s}^a)e^{2y+\frac{2}{\sqrt{7}}x}
+Ce^{ y - \sqrt{7} x  } 
-T(\tilde{s}^a)e^{\frac{3}{2} y - \frac{5}{2\sqrt{7} } x} \,,
\end{equation}
where the exponentials are multiplied by functions which correspond to the flux and source content of the theory. More specifically, we have evaluated the integral over the internal space while
considering the flux ansatzes \eqref{F4Ansatz} for the $F_4$ flux and \eqref{HAnsatz} for the $H_3$ flux; we thus find
\begin{align}
    F(\tilde{s}^a)&=\frac{1}{16} \int F_4 \wedge \tilde\star F_4 \, \text{vol}(\tilde X_7)^{\frac{1}{7}}=\frac{1}{16}\left[f^2\left(\sum_{a=1}^5 (\tilde{s}^a)^2+25(\tilde{s}^6)^2\right)+q^2\prod_{a=1}^6\frac{1}{(\tilde{s}^a)^2}\right] \label{H3}\,,\\[0.5mm]
    H(\tilde{s}^a)&=\frac{1}{16}\int H_3 \wedge \tilde \star H_3\, \text{vol}(\tilde X_7)^{-\frac{1}{7}}=\frac{h^2}{16} \sum^6_{a=1}\frac{1}{(\tilde{s}^a)^2} \,,\\[0.5mm]
    C&=\frac{m^2}{16} \,,\\[0.5mm]
    T(\tilde{s}^a)&=\frac{h m}{8} \sum_{a=1}^6 \frac{1}{\tilde s^a}\,.
\end{align}
Notice that the scalar potential in \eqref{potential} corresponds to the supergravity formula \eqref{scalarpotentialsugra}. To deduce the expression for $T(\tilde{s}^a)$ we have used the flux-charge relation from the tadpole cancellation \eqref{tad2} for a single 3-cycle to trade the charge of the source to fluxes.

Calculating then the Hessian of the potential \eqref{potential} we find the masses from the eigenvalues of the following matrix\footnote{The masses for the isotropic flux ansatz of \cite{Farakos:2020phe} are found to be
\begin{align}
    m^2L^2=\{52.625, 7.326, 3.384, 3.384, 3.384, 3.384, 3.384, 3.128\} \,, \nonumber
\end{align}
which corrects the calculation in \cite{Apers:2022zjx} where the masses are estimated to have different numerical values and hierarchy compared to this result. However, the corrected masses do not give integer dimensions for the dual operators:
\begin{align}
    \Delta=\{8.323,3.885,3.094,3.094,3.094,3.094,3.094,3.032\} \,. \nonumber
\end{align}
One arrives at the same result for the masses by using the canonical basis of the $\tilde{s}^a$'s from \cite{Andriot:2022brg}, writing the (super)potential in this basis and then calculating the eigenvalues of $\langle V_{IJ}\rangle/\vert\langle V\rangle\vert$.  
}
\begin{align}
    \text{Eigen}\left[\langle K_{IJ}\rangle^{-1}\frac{\langle V_{IJ}\rangle}{\vert\langle V\rangle\vert}\right]=m^2L^2 \,,
\end{align}
where the indices $I,J$ represent derivatives with respect to the $\tilde{s}^a$'s and the $x$, $y$ moduli. The vacuum expectation value of the potential is given by \eqref{scalarpotentialsugra} evaluated at the vacuum: $\langle V\rangle=-4P^2$. The AdS radius $L\equiv L_{\Lambda}$ is related to the vacuum expectation value of the potential via the relation
\begin{align}
    \frac{1}{L^2}=\frac{-2\langle V\rangle}{(d-1)(d-2)} \,,
\end{align}
for $d=3$. The masses for the scale-separated regime discussed in subsection \ref{scaleseparated}, were $q=f$, are found to be
\begin{align}\label{SSmasses}
    m^2L^2=\{49.778,8.178,6.347,2.589,2.589,2.589,2.589,1.966\} \,.
\end{align}
From the positive values of the actual scalar masses we can deduce the dimensions of the putative dual CFT operators. The conformal dimension $\Delta$ of an operator in a CFT is given by the standard formula in $d$-dimensions
\begin{align}
    \Delta [\Delta-(d-1)] = m^2 L^2 \,,
\end{align}
in terms of the mass $m$ and the AdS radius $L$. For the masses \eqref{SSmasses} the dimensions of the dual operators are
\begin{align}
    \Delta=\{8.126,4.029,3.710,2.894,2.894,2.894,2.894,2.722\} \,.
\end{align}
Following the same steps, we can calculate the masses of the moduli in the non-scale-separated regime, for the flux ansatzes discussed in subsection \ref{nonscaleseparated}. We summarize the results in Table \ref{CC2table}.
\begin{center}\label{table2}
\renewcommand{\arraystretch}{1.5}
\begin{tabular}{|c||c|}
     \hline 
     $c$ & $\text{Eigen}[V_{IJ}/\vert\langle V\rangle\vert]$   \\ 
     \hline \hline
     $1$   &  $\{231.942, 17.314, 3.8722, 3.4185, 0.8569, 0.8569, 0.8569, 0.8569\}$   \\
     \hline
     $10^{-1}$ &  $\{438.227, 23.672, 5.441, 3.510, 1.654, 1.654, 1.654, 1.654\}$   \\
     \hline
     $10^{-3}$  &  $\{1606.81, 65.710, 7.403, 6.167, 6.167, 6.167, 6.167, 3.523\}$  \\
     \hline
     $10^{-6}$  &  $\{11505., 436.001, 44.384, 44.384, 44.384, 44.384, 8.065, 3.525\}$  \\
     \hline
     $10^{-9}$ & $\{82741.6, 3105.15, 319.429, 319.429, 319.429, 319.429, 8.155, 3.525\}$  \\
     \hline \hline
     \multicolumn{2}{|c|}{$m^2L^2$ :  $\{49.778,8.178,6.347,2.589,2.589,2.589,2.589,1.966\}$} \\
     \hline
\end{tabular}
\captionof{table}{\label{CC2table} This table exhibits the masses of the closed string moduli as the parameter $c$ takes the values $10^{-1}$, $10^{-3}$, $10^{-6}$ and $10^{-9}$ (other than the value $c=1$). The eigenvalues of the normalized Hessian change, but, once we take into account the proper kinetic term normalization, we always find the same masses.}
\end{center}

\section{Interpolating between two flux vacua via a D4-brane} \label{D4}

In this section we will discuss how to connect the flux configuration that restores scale separation to the one that is non-scale-separated via a D4-brane. 
We will be able to describe the ``jumps'' in the parameter $Q$ from $Q = 0$ to $Q = 1$ thanks to an open string scalar associated with the D4-brane.  Furthermore, the field space distance between these two flux vacua will turn out to be finite and also realize the refined distance conjecture \cite{Ooguri:2006in,Klaewer:2017,Baume:2016}.

\begin{table}[htbp!]
\centering
\renewcommand{\arraystretch}{1.4}
\begin{tabular}{|l||c|c|c||c|c|c|c|}
\hline  & $y^2$ & $y^4$ & $y^6$ & $y^1$ & $y^3$ & $y^5$ & $y^7$ \\ 
\hline ~$F_{4,q}$  & - & - & - & $\otimes$ & $\otimes$ & $\otimes$ & $\otimes$ \\
\hline ~$H_3/\Phi_1$  & $\otimes$ & - & - & $\otimes$ & - & - & $\otimes$ \\
\hline ~$H_3/\Phi_2$  & - & $\otimes$ & - & - & $\otimes$ & - & $\otimes$ \\
\hline ~$H_3/\Phi_3$  & - & - & $\otimes$ & - & - & $\otimes$ & $\otimes$ \\
\hline ~$H_3/\Phi_4$  & - & - & $\otimes$ & $\otimes$ & $\otimes$ & - & - \\
\hline ~$H_3/\Phi_5$  & $\otimes$  & - & - & - & $\otimes$ & $\otimes$ & - \\
\hline ~$H_3/\Phi_6$  & - & $\otimes$ & - & $\otimes$ & - & $\otimes$ & - \\
\hline ~$\Phi_7$  & $\otimes$ & $\otimes$ & $\otimes$ & - & - & - & - \\
\hline
\end{tabular}
\caption{The table shows the background flux setup, where the $H_3$ flux has always only one leg on the coordinates spanned by the 3-cycle $\Phi_7$. Since the D4-brane fills AdS$_3$ and is codimension-1 on the $\Phi_7$ 
(which is Poincar\'e-dual to the $F_{4,q}$ flux), one can arrange $B_2$ to have no legs on the D4-brane world-volume.} 
\label{H3F4-table} 
\end{table}

\subsection{The D4-brane Bianchi identity}

Following \cite{Shiu:2022oti}, in order to interpolate between different $F_{4,q}$ flux numbers we can introduce a D4-brane that fills the three-dimensional AdS$_3$ spacetime and wraps a contractible 2-cycle within a 3-cycle in the internal seven-dimensional space. In particular, we would like such D4-brane to be a codimension-1 object inside the 3-cycle $\Sigma_{3,7} \equiv \Phi_7$ that is Poincair\'e dual to the 4-cycle $\Sigma_{4,7} \equiv \Psi_7$ filled with the $F_4$ flux, according to \eqref{F4Expansion}.
In the presence of a D4-brane, the $F_4$ Bianchi identity gives
\be
\text{d}F_4 = (2\pi)^7 Q_{\rm D4} \, \delta(\psi-\psi_0,y^1,y^3,y^5,y^7) \, \text{d}\psi \wedge \epsilon_4 \,, 
\ee
where $\psi$ and $\{y^i\}_{i=1,3,5,7}$ denote the internal coordinates corresponding to the transverse directions to the D4-brane, inside $\Sigma_{3,7}$ and for $\Sigma_{4,7}$ respectively; $\psi_0$ is the actual position of the D4-brane along the coordinate $\psi$ and $\epsilon_4$ represents the normalized volume 4-form of $\Sigma_{4,7}$, i.e. $\epsilon_4 = \frac{r_1 \text{d}y^1 \wedge r_3 \text{d}y^3 \wedge r_5 \text{d}y^5 \wedge r_7 \text{d}y^7}{r_1 r_3 r_5 r_7} = \Psi_7$. If we smear the D4-brane over the 4-cycle $\Sigma_{4,7}$, the $F_4$ form Bianchi identity becomes
\be
\text{d}F_4 = (2\pi)^7 Q_{\rm D4} \, \delta(\psi-\psi_0) \, \text{d}\psi \wedge \Psi_7 \,, 
\ee
so that, when moving between the two sides ($\psi<\psi_0$ and $\psi>\psi_0$) of the D4-brane, the $F_4$ flux changes by
\be\label{DeltaF4}
\Delta F_4 = (2\pi)^7 Q_{\rm D4} \, \theta(\psi-\psi_0) \Big{\vert}_{\psi<\psi_0}^{\psi>\psi_0} \Psi_7 = (2\pi)^7 Q_{\rm D4} \, \Psi_7 \,,
\ee
where $\theta(\cdot)$ is the Heaviside step function. We can exploit the emergence and pinching off of such D4-brane to pass (non-dynamically) from one flux vacuum to another one (see Figure 1 and Figure 2). More precisely, to transit from the regime where scale separation is broken at $Q=0$ (or $S=-\frac17$) to the configuration where there is scale separation at $Q=1$ (or $S=0$) we need to account for a change of $N+1$ units in the $F_{4,q}$ flux. We can then repeat the process multiple times and produce the desired change of $F_{4,q}$ flux units, after the D4-branes, wrapped over a 2-cycle, have swept out the whole 3-cycle $\Sigma_{3,7}$: the charge of a D4-brane being $Q_{\rm D4} = \frac{1}{(2\pi)^4}$, scanning $n$ times the 3-cycle $\Sigma_{3,7}$, the $F_{4,q}$ flux changes by $n$ units according to \eqref{DeltaF4}, i.e. $\Delta f_{4,q} = (2\pi)^3 n$, as we would like to achieve.

\subsection{The EFT of the D4-brane}

Let us consider a probe D4-brane within our ten-dimensional spacetime $\text{AdS}_3 \times X_7$. We would like to show that this brane, other than extending over the external three-dimensional AdS$_3$ space, can wrap, after a reparametrization, a contractible 2-cycle within the 3-cycle $\Sigma_{3,7} \equiv \Phi_7$, which involves the internal directions $y_2$, $y_4$ and $y_6$. Furthermore, we would infer that the scalar field, say $\psi$, which is associated to the transverse direction to the brane inside $\Sigma_{3,7}$ and parametrizes the distance travelled when the $F_{4,q}$ flux changes, belongs to the effective field theory description we rely on. To this specific purpose we are going to compare the energy density of the D4-brane to the vacuum energy density.

For convenience the calculation will be performed by introducing coordinates adapted to the lower half of the 3-cycle of Figure 1, 
therefore the coordinate $y_2$ has half range compared to the others. 
We can trade the parametrization in \eqref{lineelement} for
\be
\text{d}\tilde{s}^2_{7} = \text{d} \tilde{s}^2_{\Sigma_{3,7}} + \text{d} \tilde{s}^2_{\Sigma_{4,7}} = \left( \tilde{r}_2^2 \text{d} y_2^2 +  \tilde{r}_4^2 \text{d} y_4^2 + \tilde{r}_6^2 \text{d} y_6^2 \right) + \text{d}\tilde{s}^2_{\Sigma_{4,7}}
\ee
(where $\Sigma_{4,7} \equiv \Psi_7$ is the Poicair\'e dual cycle to $\Phi_7$ and $y_2$ ranges from $0$ to $\frac12$) and consequently
\be \label{IntMetric}
\text{d}\tilde{s}^2_{7} = \tilde{s}_7^{\frac23} \left( \tilde{\rho}_1^2 \text{d}\tilde{y}_2^2 +  \text{d}\tilde{y}_4^2 + \tilde{\rho}_2^2 \text{d}\tilde{y}_6^2 \right) + \text{d}\tilde{s}^2_{\Sigma_{4,7}} \,.
\ee
Let us note that the values of the moduli $\tilde{r}_i$ are fixed in the vacuum configurations of our interest. 
In particular, we consider the moduli to be fixed to their values at the initial vacuum configuration, say the one that breaks scale separation (at $S \lesssim -\frac17$). 
Then, once the D4-brane contracts as the other vacuum configuration is reached (at $S=0$), we imagine to un-freeze the moduli and let them take their new critical values. 

This being clarified, we decide to rewrite the ($\tilde{y}_2,\tilde{y}_4,\tilde{y}_6$)-sector of \eqref{IntMetric} by introducing spherical coordinates, i.e.
\be
\text{d}\tilde{s}^2_{7} = \tilde{s}_7^{\frac23} \left[ \text{d} R^2 +  R^2 \left( \text{d}\psi^2 + (\sin\psi)^2 \text{d}\omega^2\right) \right] + \text{d}\tilde{s}^2_{\Sigma_{4,7}} \, ,
\ee
where, because of the structure of \eqref{IntMetric}, the coordinates $R$, $\psi$ and $\omega$ need to satisfy intricate boundary conditions
\be \label{BoundaryConditions}
\mathcal{F}_k (R,\psi,\omega) = 0 \quad\quad \text{with} \quad k=1,2,\dots,6
\ee
(one for each of the faces of the parallelepiped $\Sigma_{3,7}$). The D4-brane wraps the radial $R$ and the angular $\omega$ directions; the angle $\psi$ parametrizes, instead, its movement within the 3-cycle and makes the interpolation between the two flux vacua of interest achievable (see Figure 2). 
We then make the coordinate $\psi$ to correspond to one of the five scalar fields that give the position of the D4-brane over spacetime via a localization in the five transverse directions to its world-volume. 
In order not to spread our notation much, we call this scalar field again $\psi$, which travels from $\psi=0$ to $\psi=\pi$ when reducing the $F_{4,q}$ flux by $(2 \pi)^3$. This means that, when we change the flux by $n \times (2 \pi)^3$, the scalar movement corresponds to a repetitive and consecutive movement of $n$ scalars corresponding to $n$ D4-branes (which should not be considered as a stack). It is as if the scalar $\psi$ starts from taking value $0$ and, after $n$ consecutive steps takes the value $n \times \pi$. This will be relevant later on for the evaluation of the distance.  
Once the field $\psi$ is allowed to depend on the external coordinates and a pull-back of 
\be
\text{d}s_{10}^2 = e^{2\alpha v} \text{d}s_{3}^2 + s_7^{\frac23} \left[ \text{d} R^2 + R^2 \left( \text{d}\psi^2 + (\sin\psi)^2 \text{d}\omega^2\right) \right] + e^{2\beta v} \text{d}\tilde{s}^2_{\Sigma_{4,7}} 
\ee
onto the brane is performed, we can extrapolate
\be \label{pulledD4metric}
\text{d}s^2_{\text{D4}} = \left( e^{2\alpha v} g^{(3)}_{\mu \nu} + s_7^{\frac23} R^2 \partial_{\mu} \psi \partial_{\nu} \psi \right) \text{d}x^{\mu} \text{d}x^{\nu} + s_7^{\frac23} \left( \text{d} R^2 + R^2 (\sin\psi)^2 \text{d}\omega^2 \right) \,,
\ee
$g^{(3)}_{\mu \nu}$ being the three-dimensional AdS$_3$ metric. The bosonic sector of a D4-brane action in the Einstein frame is
\be \label{D4action}
S_{\text{D4}} = -T_4 \int \text{d}^5 x \, e^{\frac{\phi}{4}} \sqrt{-\text{det}[ g^{(5)} + 2\pi \cal{F}]} + \text{Chern--Simons term} \,,
\ee
where $T_4$ is the tension coefficient of the D4-brane, $T_4 = \frac{1}{(2\pi)^4}$ (in string units); $g^{(5)}$ stands for the metric associated to \eqref{pulledD4metric}, and $\cal{F}$ is the 2-form given by the $B_2$ NSNS field and the field strength $F$ of the 1-form $A_1$ leaving on the brane world-volume.

\begin{figure}[t]
\includegraphics[scale=0.19]{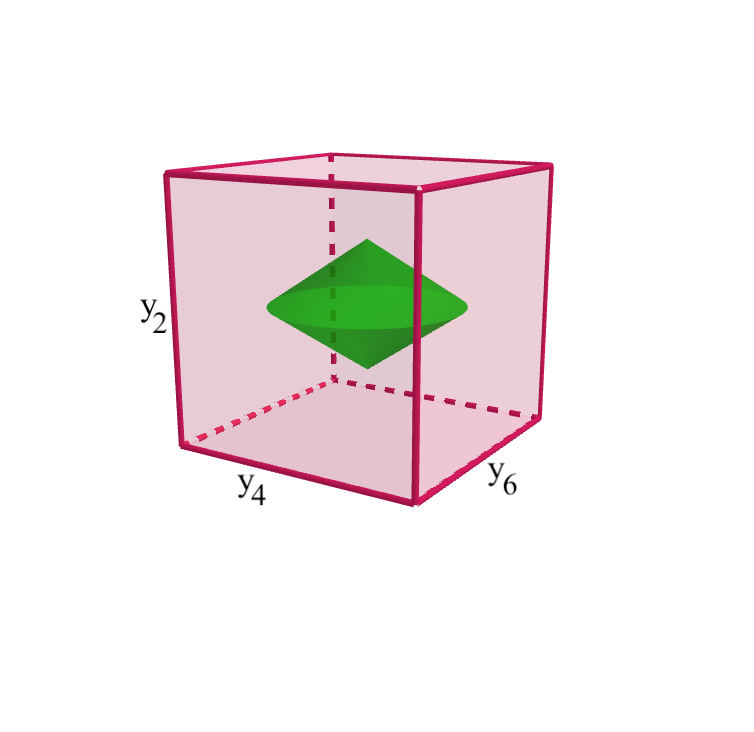}
\includegraphics[scale=0.19]{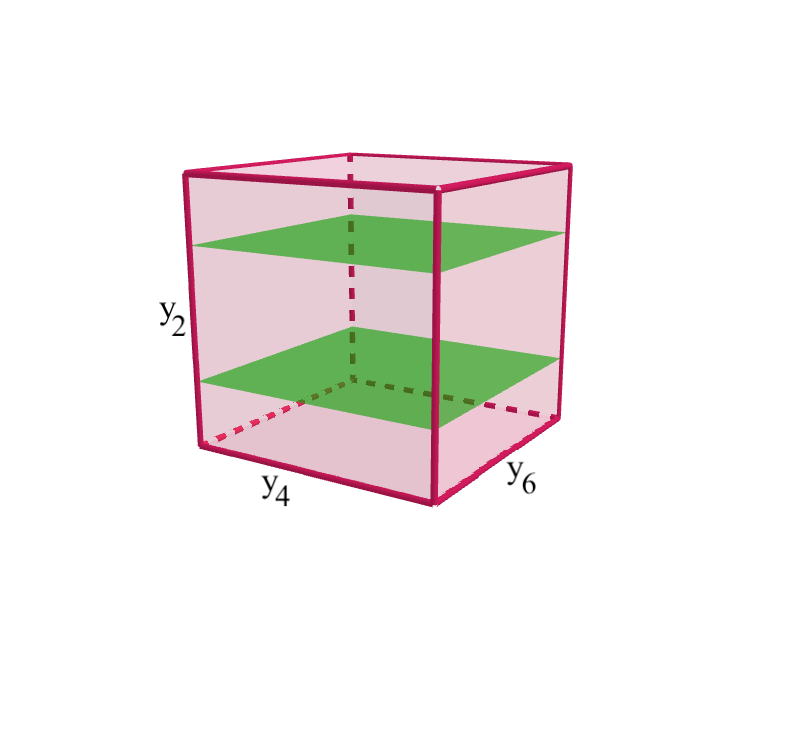}
\includegraphics[scale=0.19]{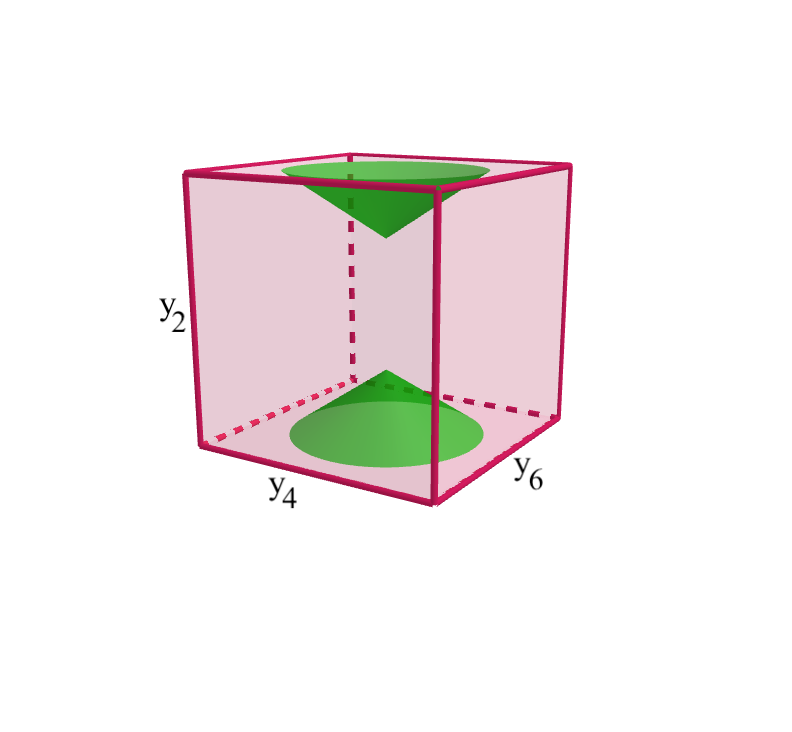}
\label{Full-Torus}
\centering
\caption{The figure shows the closed surface wrapped by the D4-brane inside the full 3-torus of the internal space at three instances instances: 
$\psi_{\rm left} = \pi/4$, $\psi_{\rm center} = \pi/2$ and $\psi_{\rm right} = 3 \pi /4$, where $\psi$ measures the angle of the lower half of the cycle. 
In our calculations and related plots we focus on the lower half of the present figure: the upper half is just a mirror of the former. Moreover, despite the fact that in a realistic construction all vertices or edges should be smooth, our upcoming calculation of the field space distance (for example) is done with sharp edges and vertices, which should be a quite good approximation.}
\end{figure}

\begin{figure}[t]
\includegraphics[scale=0.35]{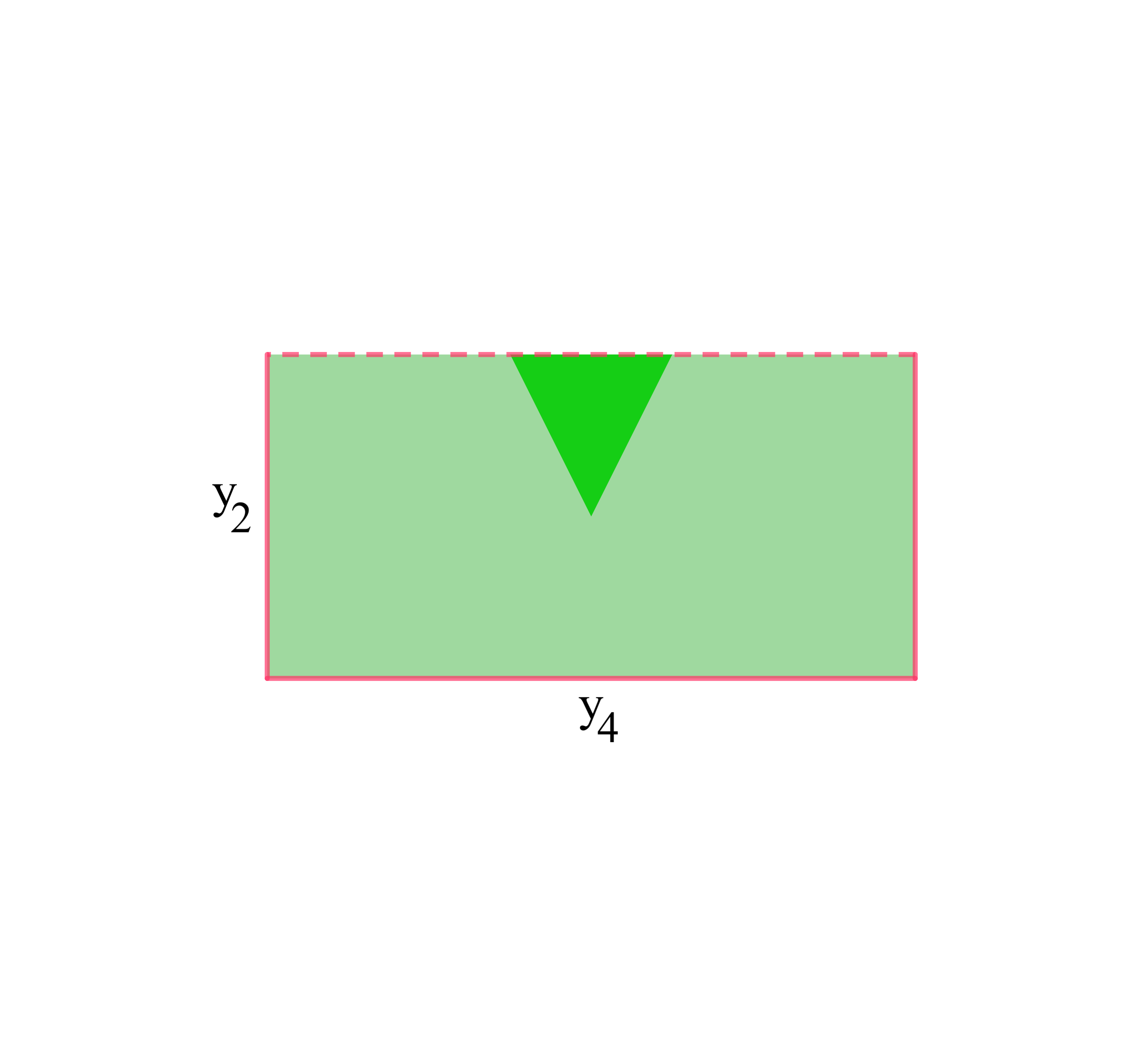}
\includegraphics[scale=0.35]{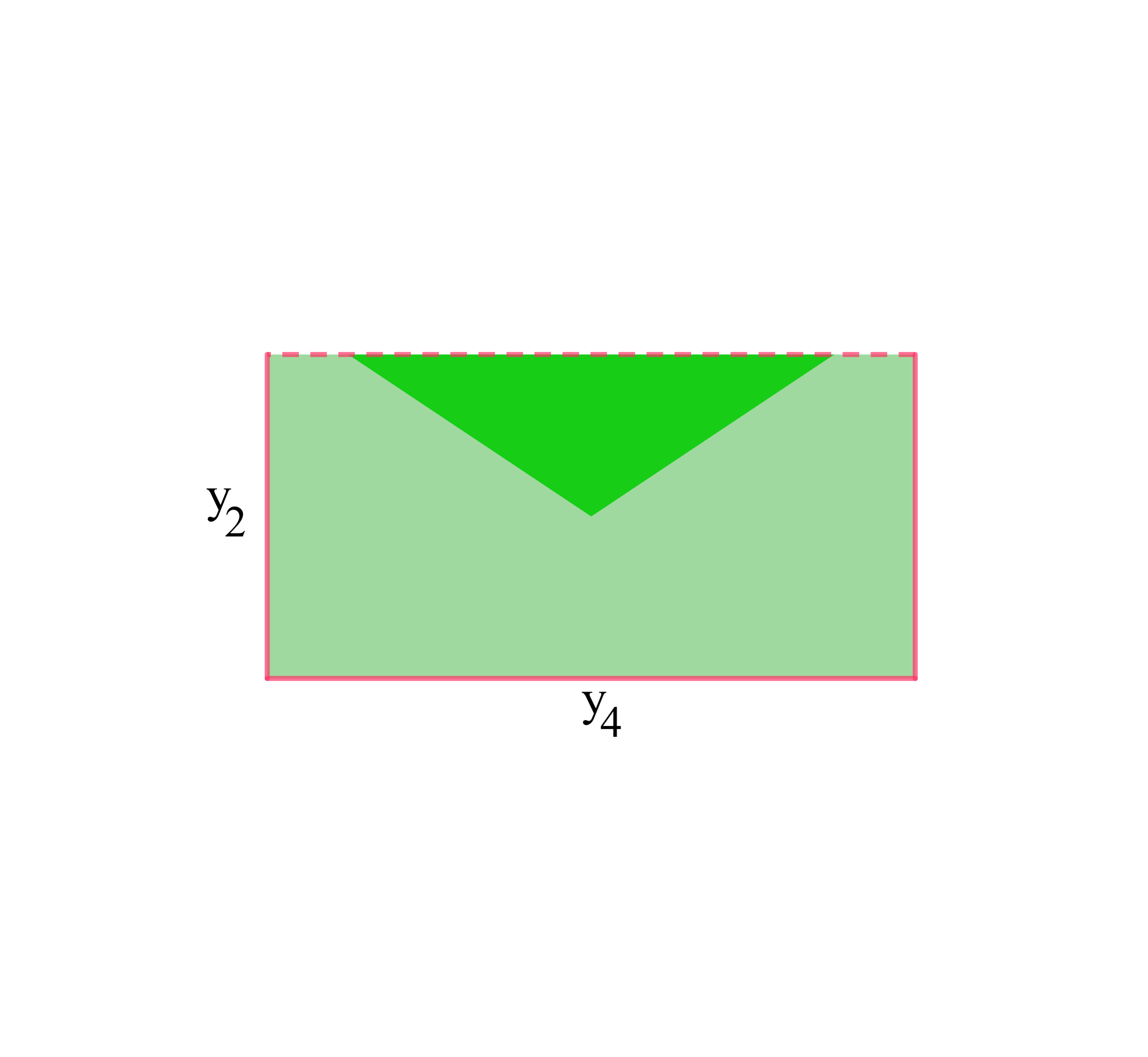}
\includegraphics[scale=0.35]{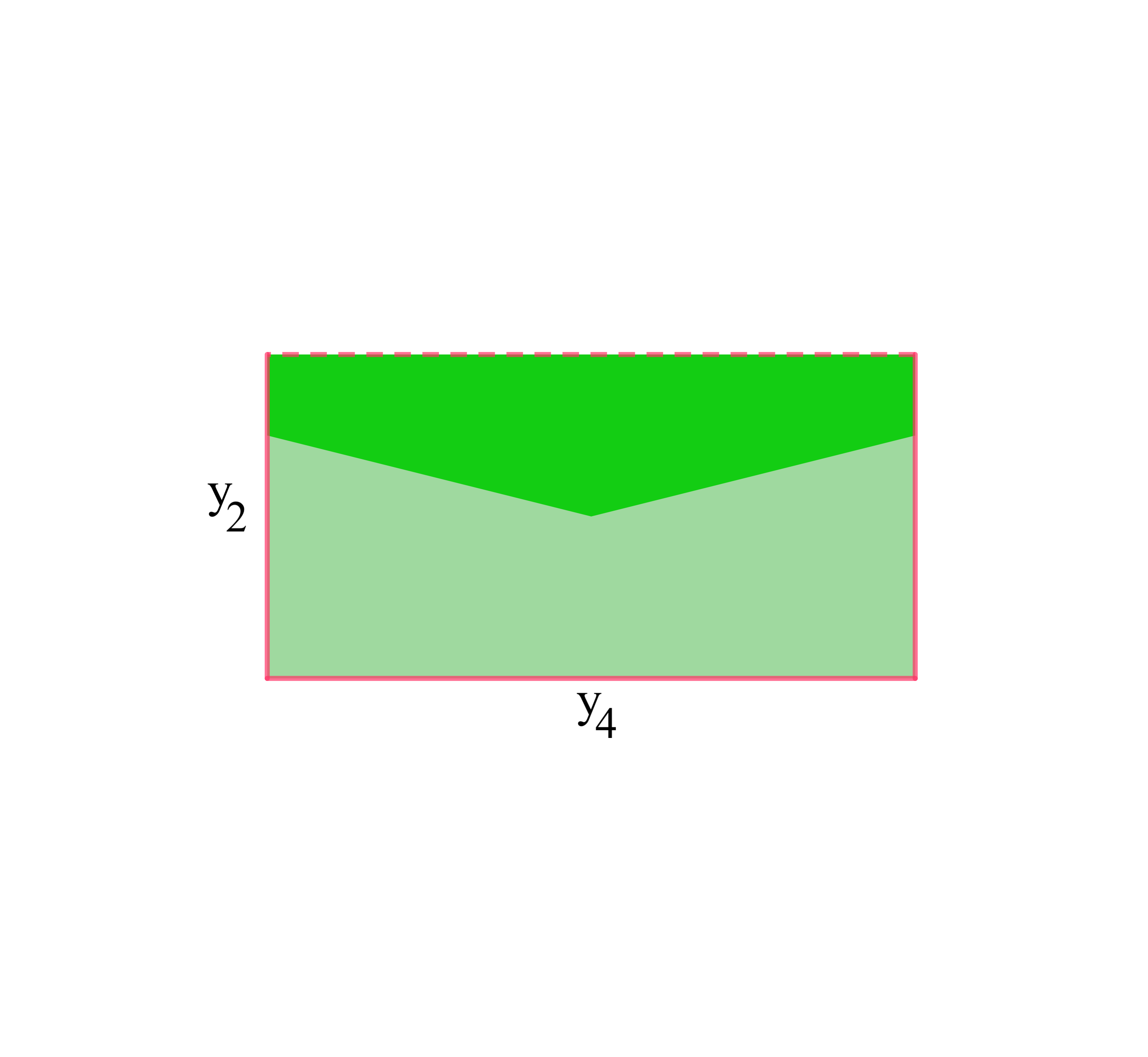}
\includegraphics[scale=0.35]{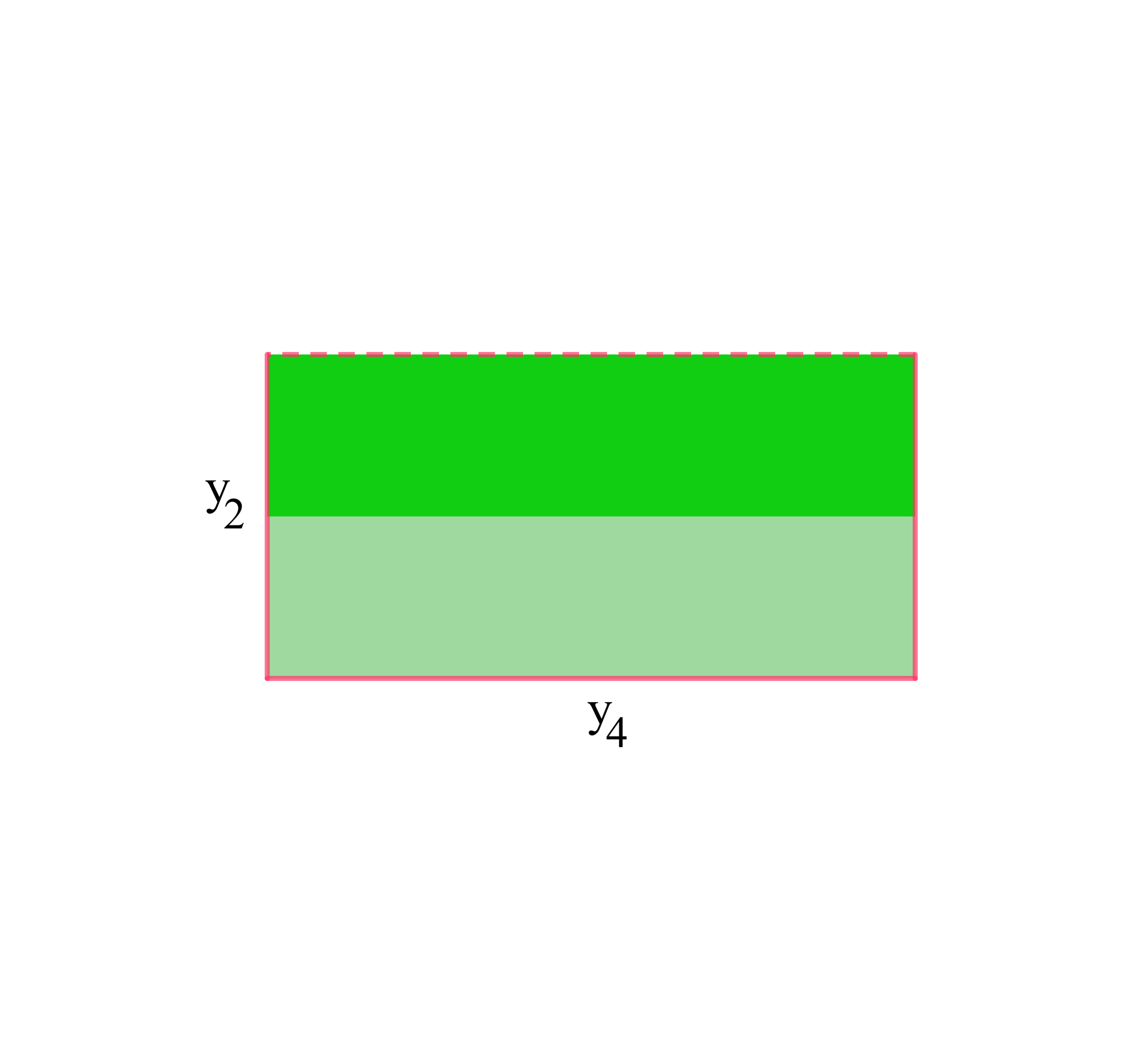}
\includegraphics[scale=0.35]{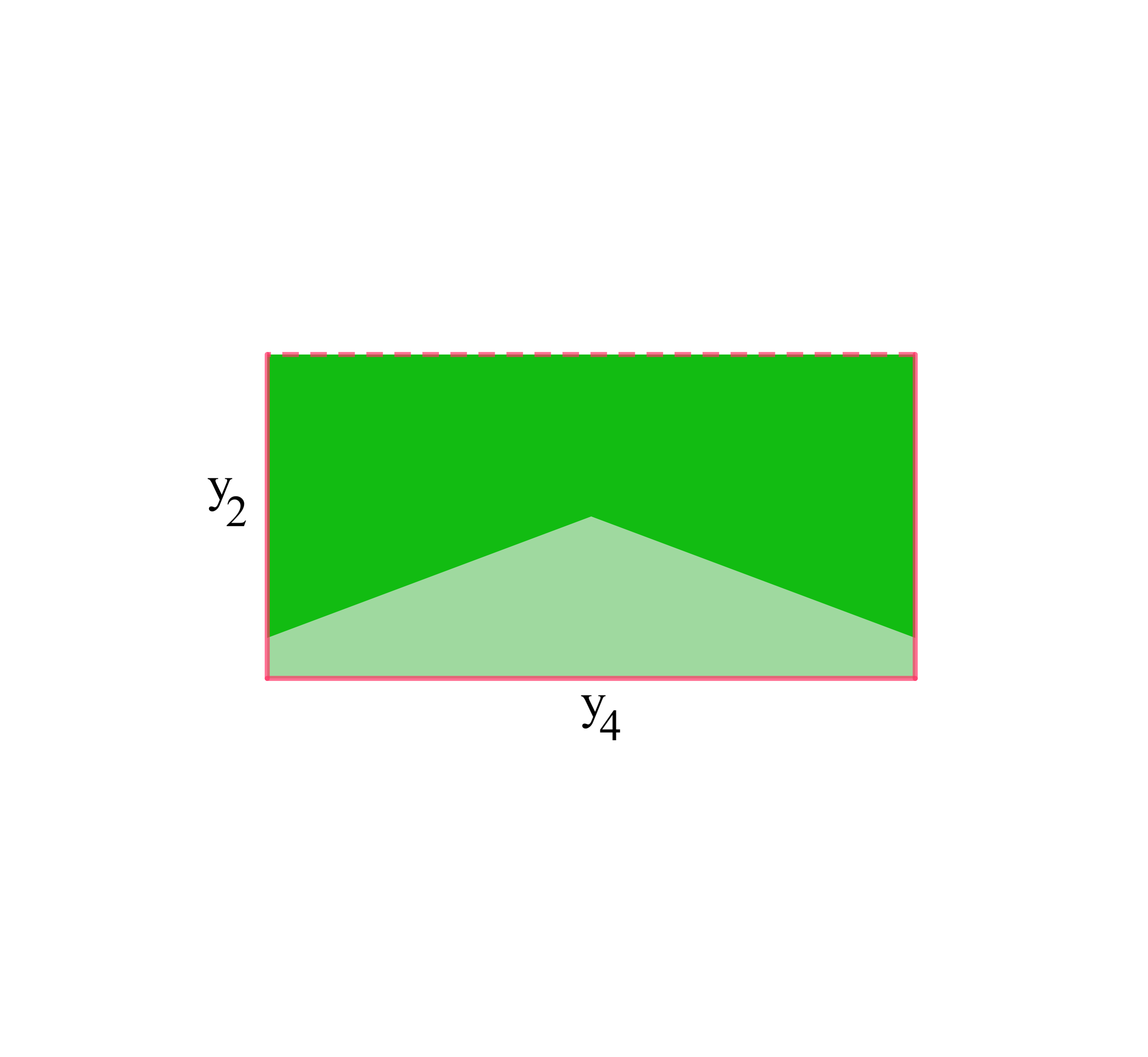}
\includegraphics[scale=0.35]{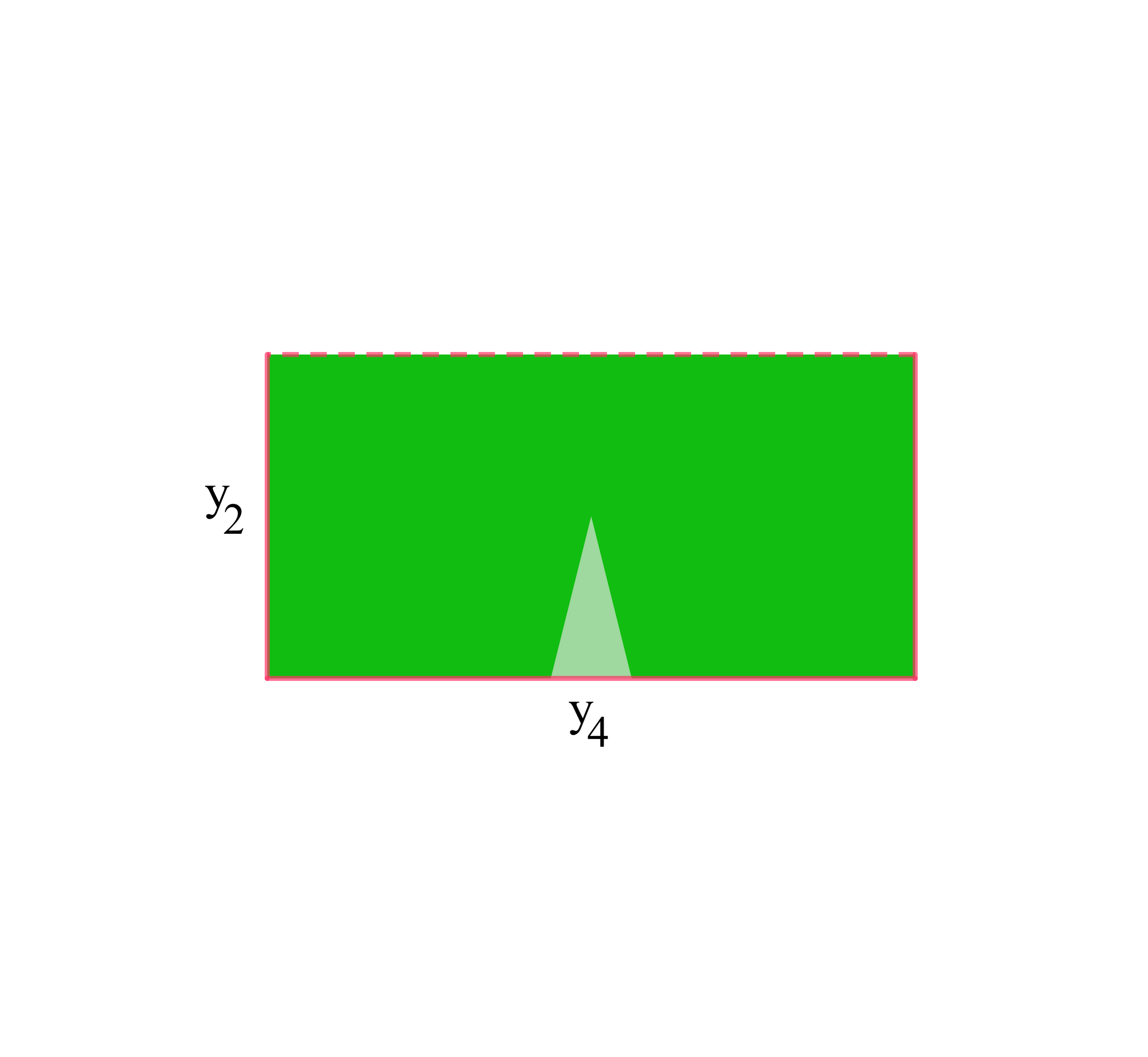}
\label{D4brane wrapping}
\centering
\caption{The figure schematically shows a two-dimensional vertical slice of the lower half of the 3-cycle $\Sigma_{3,7}$ of Figure 1, representing a single D4-brane configuration that interpolates between two vacua with different ($q$ and $q-(2\pi)^3$) flux units along $\Psi_7$. The D4-brane wraps AdS$_3$ and a contractible 2-cycle inside a 3-cycle spanned by $y_2$, $y_4$ and $y_6$. While moving between the two configurations, the D4-brane scans the whole 3-cycle. The six D4-brane positions that the reader can see in the figure correspond to different instances of the D4-brane modulus $\psi$ in the interval $[0,\pi]$. The darker and lighter green regions correspond to different $F_{4,q}$ flux values, $q$ and $q-(2\pi)^3$ respectively.} 
\end{figure}

Because of our ansatz for the $H_3$ flux, see \eqref{HAnsatz}, we can consistently choose the $B_2$ field profile so that it does not have any leg on the directions of $\Sigma_{3,7}$, namely the ones corresponding to $\{y^m\}_{m=2,4,6}$ (in the Cartesian basis; see, in particular, Table \ref{H3F4-table}). The only non-vanishing components of $B_2$ are then
\begin{align}
    [B_2]_{17}(y^2)\,,\quad [B_2]_{37}(y^4)\,,\quad [B_2]_{57}(y^6)\,,\quad [B_2]_{13}(y^6)\,,\quad [B_2]_{35}(y^2)\,,\quad [B_2]_{15}(y^4) \,,
\end{align}
and they satisfy 
\begin{align}
    \int_{\Sigma_{3,i}} \text{d}B_{2,i} = h  \quad \text{(for }  i \ne 7 \text{ )} \quad\quad \text{and} \quad\quad \int_{\Sigma_{3,7}} \text{d}B_{2,7} = 0 \,, 
\end{align}
for each 3-cycle $\Sigma_{3,i}$. As a consequence, we can forget about $\cal{F}$ in \eqref{D4action}. 
Moreover, because $B_2$ never intersects with the D4-brane world-volume, we do not get any effective net D2-charge (which is a coupling to the RR 3-form $C_3$) from the Chern--Simons term of \eqref{D4action}, in contrast to \cite{Shiu:2022oti}. As far as our purpose is concerned, we can consistently look only at the resulting DBI piece of $S_{\text{D4}}$\footnote{If the Chern--Simons term is also included, the following discussion does not qualitatively change \cite{Shiu:2022oti}; we also do a crude estimate of the scaling of the Chern--Simons piece in Appendix \ref{ChernSimons} and see that it is subdominant.}, i.e.
\be
S_{\text{D4}} = -T_4 \int \text{d}^5 x \,e^{\frac{\phi}{4}} \sqrt{-\text{det}[ g^{(5)} ]} + \dots .
\ee
Once evaluated on \eqref{pulledD4metric} it becomes
\be \label{pulledD4action}
\begin{aligned}
S_{\text{D4}} = -\frac{(2\pi)^{21} T_4}{8} \int \text{d}^3x & \sqrt{-\text{det}[g^{(3)}]} \,  e^{\frac{\phi}{4}} e^{3\alpha v} s_7^{\frac23} \times \\ & \times \int_{R_{-}}^{R_{+}} \text{d}R \int_{\omega_{-}}^{\omega_{+}} \text{d}\omega \, R \sin\psi \, \sqrt{1+ \frac{4 s_7^{\frac23} e^{-2\alpha v} R^2}{(2\pi)^{14}} (\partial \psi)^2} \, + \dots \,,
\end{aligned}
\ee
once the Weyl rescaling of the three-dimensional metric $g^{(3)}_{\mu \nu} \to \frac{(2\pi)^{14}}{4} g^{(3)}_{\mu \nu}$ has been performed, and $R_{\pm}$ and $\omega_{\pm}$ are the boundary values of the radial and angular coordinates $R$ and $\omega$ (respectively), dictated by the structure of $\Sigma_{3,7}$.
From \eqref{pulledD4action} we can extract the scalar potential
\be
V(\psi) \supset \frac{(2\pi)^{21} T_4 \, u}{8} \, e^{ \frac{\phi}{4} - 21\beta v} s_7^{\frac23}  \sin\psi \,,
\ee
where the parameter $\alpha$ is $\alpha=-7\beta$ (see Section \ref{Setup}) and the constant $u$ is given by $u = \int_{R_{-}}^{R_{+}} \text{d}R \int_{\omega_{-}}^{\omega_{+}} \text{d}\omega \, R$. Because of its dependence on $e^{\phi}$ (or, alternatively, $g_s$), $e^{\beta v}$ and $s_7 = r_2 r_4 r_6$, by referring to \eqref{gsscaling}, \eqref{vscaling} and \eqref{r246scaling} the energy density of the D4-branes scales with $N$ as
\be
V(\psi) \sim N^{-\frac{3 + 7S}{16}} \times N^{-\frac{21}{16}(7 + 11S)} \times N^{\frac{7 + 11S}{8} - 4S} = N^{-\frac{17 + 35 S}{2}} \,.
\ee

As mentioned above, in order to argue for the consistency of the inclusion of $\psi$ in the effective description we would like to rely on, $V(\psi)$ has to be compared to the vacuum energy density, which we denote as $V_{\text{flux}}$. Since we are working in $\alpha^\prime$ units, its scaling with $N$ can be deduced simply by the one of $P^2$, which is $P^2 \sim N^{-8 - 14S}$. 
Let us then observe that
\be
\frac{V(\psi)}{|V_{\text{flux}}|} \sim N^{-\frac{1 + 7S}{2}}
\ee
and, interestingly,
\be\label{energyscaling}
\frac{V(\psi)}{|V_{\text{flux}}|} \sim N^{0} \quad \text{for }  S=-\frac17 \quad\quad \text{and} \quad\quad \frac{V(\psi)}{|V_{\text{flux}}|} \sim N^{-\frac{1}{2}} \quad \text{for }  S=0 \,. 
\ee
In the flux vacuum where scale separation is broken the energy density of the D4-brane and the absolute value of the vacuum energy density tend to be of the same order of magnitude, even though, if we take $\epsilon \text{ (}= -1-7S$) into account the D4-brane potential can always be parametrically smaller than the vacuum energy density. 
Besides this, in the configuration where scale separation is realized the energy density of the D4-brane is directly lower than the absolute value of the vacuum energy density. 

From \eqref{pulledD4action} we can also deduce the metric component $g_{\psi \psi}$, which accompanies the kinetic term for the canonically normalized field $\psi$ in the effective theory of the D4-brane. As we are going to see soon, $g_{\psi \psi}$ represents a crucial ingredient for the application of the distance criterion. Precisely, it is
\be
g_{\psi\psi} = \frac{(2\pi)^7 T_4 \, k}{2} \, e^{ \frac{\phi}{4} - 7\beta v} \, s_7^{\frac43} \sin\psi \,,
\ee
the constant $k$ being $k = \int_{R_{-}}^{R_{+}} \text{d}R \int_{\omega_{-}}^{\omega_{+}} \text{d}\omega \, R^3$. The actual quantity that takes part to the expression of the geodesic distance in \eqref{DCformula} is $g_{\psi\psi}/M_P$, whose scaling with $N$ has to be checked. We have
\be
\frac{g_{\psi\psi}}{M_P} \sim N^{\frac{7 + 11S}{4} - 8S - \frac{3 + 7S}{16} - \frac{7}{16}(7 + 11S)} = N^{-\frac{3 + 21 S}{2}} \,,
\ee
which becomes
\be\label{metricscaling}
\frac{g_{\psi\psi}}{M_P} \sim N^{0} \quad \text{for }  S=-\frac17 \quad\quad \text{and} \quad\quad \frac{g_{\psi\psi}}{M_P} \sim N^{-\frac{3}{2}} \quad \text{for } S=0 \,.
\ee
We would like the reader to notice that the scalings with $N$ in \eqref{energyscaling} and \eqref{metricscaling} for the $S = 0$ case agree with \cite{Shiu:2022oti}.

\subsection{The distance conjecture}

To close this chapter we would like to evaluate the distance between the flux vacuum where scale separation is broken (i.e. for $S=-\frac17$ or $Q=0$) and the flux vacuum where, instead, scale separation is realized (i.e. for $S=0$ or $Q=1$). 
We will firstly show that the distance is finite and we will then elaborate on the distance conjecture using the geodesic path.

Clearly, AdS vacua with scale separation are in tension with the strong version of the so-called AdS distance conjecture \cite{Lust:2019zwm}, 
as they would be at infinite distance. 
However, the conjecture \cite{Lust:2019zwm} relies on an extension of the standard distance conjecture (i.e. \cite{Ooguri:2006in}) which instead we employ here. 
The later is satisfied by our analysis, for the set of scalar fields that we consider. 
Nevertheless, a weaker version of the AdS distance conjecture that requires the KK modes to drop as $m_{\text{KK}} \sim |V|^\alpha$ for $\alpha$ positive 
in the limit $V \to 0$ 
is satisfied by the scale-separated vacua ($S=0$) since as we have seen for large $N$ the vacuum energy goes to zero as $V \sim P^2 \sim N^{-8}$ 
and the same happens also for the KK masses which scale as $m_{\text{KK}}^2 \sim e^{-16 \beta v} \sim N^{-7}$ which also goes to zero for large $N$.

\subsubsection{The field space distance}

We will now deduce the expression for the distance and then show that it is finite. 
In order for this purpose to be achieved we need to know the scalar field space metric components. As we already deduced, the first relevant ingredient is the metric element $g_{\psi\psi}$ that, once normalized by the proper power of $M_P$, is
\be
\begin{aligned}
\frac{g_{\psi\psi}}{M_P} = \frac{(2\pi)^4 T_4 \, k}{2\sqrt{\pi}} \, e^{\frac{\phi}{4} - 7\beta v} s_7^{\frac43} \, \sin\psi \,= 2 c \, k \, \tilde{s}_7^{\frac43} \, e^{\frac{\phi}{4} - 3\beta v} \sin\psi \,
\end{aligned}
\ee
($c$ being a constant). 
Other than $g_{\psi\psi}$ we also need the metric components $g_{\phi\phi}$, $g_{v v}$ and $g_{\tilde{s}^a \tilde{s}^b} $, which can be extracted from the dimensionally reduced ten-dimensional action over the $G_2$ space, once the fields $\phi$, $v$ and $\tilde{s}^a$'s have been canonically normalized. 
In particular, 
\be
g_{\phi \phi} = \frac12 M_P \,, \quad\quad g_{v v} = \frac12 M_P \quad\quad \text{and} \quad\quad g_{\tilde{s}^a \tilde{s}^b} = \frac{1+\delta_{a b}}{2\tilde{s}^a \tilde{s}^b} \quad \text{for } \  a, b = 1,...,6 \,.
\ee
The field space distance we are interested in is
\be
\Delta = \int_{\xi=0}^{\xi=1} \text{d}\xi \ \sqrt{\frac{g_{A B}}{M_P} \frac{\text{d}\varphi^A}{\text{d}\xi} \frac{\text{d}\varphi^B}{\text{d}\xi}} \,,
\ee
where the indices $A$ and $B$ run over the scalar fields $\varphi^A = \{\psi, \phi, v, \tilde{s}^a\}$ and $g_{A B}$ denote the corresponding metric elements. By exploiting reparametrization invariance we can impose that $\xi=0$ corresponds to the flux vacuum with 1 unit of $F_{4,q}$ flux and $\xi=1$ to the vacuum configuration with, say, $N$ units of $F_{4,q}$ flux, so that as $\xi$ changes from $0$ to $1$ we approach the restoration of scale separation from its breaking regime.

We would like to show that the flux vacua at $\xi=0$ and $\xi=1$ are at finite distance one with respect to the other. To this purpose it is enough to identify a path for which $\Delta$ is finite: in fact, as a consequence of its definition, also the geodesic distance between the flux vacua of interest would then be finite. We have
\be
\frac{\Delta}{\sqrt{2}} = \int_0^1 \text{d}\xi \left[ \frac{c \,  k \, e^{\frac{\phi}{4} - 3\beta v} \sin\psi}{\left(\prod_{a=1}^6\tilde{s}^a\right)^{\frac43}} \left(\frac{\text{d}\psi}{\text{d}\xi}\right)^2 + \frac14 \left(\frac{\text{d}\phi}{\text{d}\xi}\right)^2 + \frac14 \left(\frac{\text{d} v}{\text{d}\xi}\right)^2 + \sum_{a,b=1}^6 \frac{1+\delta_{a b}}{4\tilde{s}^a \tilde{s}^b} \frac{\text{d}\tilde{s}^a}{\text{d}\xi} \frac{\text{d}\tilde{s}^b}{\text{d}\xi} \right]^{\frac12} . \,\,
\ee
The distance becomes
\be
\begin{aligned}
\frac{\Delta}{\sqrt{2}} = \int_0^1 \text{d}\xi \Bigg[ \frac{e^{\frac{\phi}{4} - 3\beta v }}{(\sigma^5\tau)^{\frac43}} \left(\frac{\text{d} \chi}{\text{d}\xi}\right)^2 +& \frac14 \left(\frac{\text{d}\phi}{\text{d}\xi}\right)^2 + \frac14 \left(\frac{\text{d} v}{\text{d}\xi}\right)^2 + \\ & + \frac{15}{2\sigma^2} \left(\frac{\text{d} \sigma}{\text{d}\xi}\right)^2 + \frac{5}{2\sigma \tau} \frac{\text{d} \sigma}{\text{d} \xi}\frac{\text{d} \tau}{\text{d} \xi} + \frac{1}{2\tau^2} \left(\frac{\text{d} \tau}{\text{d}\xi}\right)^2 \Bigg]^{\frac12} \,,
\end{aligned}
\ee
once the variable $\chi$ has been introduced via
\be
\frac{\text{d} \chi}{\text{d}\psi} = \sqrt{c k \sin\psi} \,,
\ee
and the ansatz \eqref{SSRAnsatz} for the moduli fields has been used.
Moreover, after $\sigma$ and $\tau$ have been redefined as
\be
\sigma = e^{3\theta} \quad\quad \text{and} \quad\quad \tau = e^{3\eta} \,,
\ee
$\Delta$ takes the form
\be\label{MasterDelta}
\begin{aligned}
\frac{\Delta}{\sqrt{2}} = \int_0^1 \text{d}\xi \Bigg[ e^{\frac{\phi}{4} - 3\beta v -20\theta - 4\eta} \left(\frac{\text{d} \chi}{\text{d}\xi}\right)^2 +& \frac14 \left(\frac{\text{d}\phi}{\text{d}\xi}\right)^2 + \frac14 \left(\frac{\text{d} v}{\text{d}\xi}\right)^2 + \\ & + \frac{135}{2} \left(\frac{\text{d} \theta}{\text{d}\xi}\right)^2 + \frac{45}{2} \frac{\text{d} \theta}{\text{d} \xi}\frac{\text{d} \eta}{\text{d} \xi} + \frac92 \left(\frac{\text{d} \eta}{\text{d}\xi}\right)^2 \Bigg]^{\frac12} \,.
\end{aligned}
\ee

Before going to the more intricate evaluation of the distance \eqref{MasterDelta} with the use of the geodesic path, 
we can readily show that the distance $\Delta$ is finite. 
We can, for example, evaluate $\Delta$ along the path 
\be \label{path}
 \Gamma : \left\{
    \begin{matrix}
        \chi =  a_1 \xi + a_0  \\[3mm]
        \phi =  b_1 \xi + b_0  \\[3mm]
        v = c_1 \xi + c_0  \\[3mm]
        \theta = d_1 \xi + d_0  \\[3mm] 
        \eta = e_1 \xi + e_0  
    \end{matrix}
\right.  \,,
\ee
where $\{a_j\}_{j=0,1}$, $\{b_j\}_{j=0,1}$, $\{c_j\}_{j=0,1}$, $\{d_j\}_{j=0,1}$ and $\{e_j\}_{j=0,1}$ are some real coefficients that are fixed by the boundary conditions of the path. We thus obtain
\be \label{non-geodesic distance}
\begin{aligned}
& \Delta_{\Gamma} = \int_0^1 \text{d}\xi \sqrt{2 a_1^2 \, e^{A\xi + B} + 2 C} = \frac{\sqrt{8 C}}{A} \left( \sqrt{1 + \frac{a_1}{C} e^{A \xi +B}} - \text{arctgh}\left[ \sqrt{1+\frac{a_1}{C}e^{A \xi +B}} \right] \right) \Big{|}_{\xi=0}^{\xi=1} = \\[1mm] & = \frac{\sqrt{8 C}}{A} \left( \sqrt{1 + \frac{a_1}{C} e^{A+B}} - \text{arctgh}\left[ \sqrt{1+\frac{a_1}{C}e^{A+B}} \right] - \sqrt{1 + \frac{a_1}{C} e^{B}} + \text{arctgh}\left[ \sqrt{1+\frac{a_1}{C}e^{B}} \right] \right) \,,
\end{aligned}
\ee
with
\begin{align}
A &= \frac{b_1}{4} - 3\beta c_1 -20 d_1 - 4 e_1 \,, \\[1mm]
B &= \frac{b_0}{4} - 3\beta c_0 -20 d_0 - 4 e_0 \,, \\[1mm]
C &= \frac14 b_1^2 + \frac14 c_1^2 + \frac{135}{2} d_1^2 + \frac{45}{2} d_1 e_1 + \frac92 e_1^2 \,. 
\end{align}
The distance $\Delta_{\Gamma}$ evaluated over the path \eqref{path} is finite.
This consequently proves that \textit{causis maioribus} the geodesic distance between the flux vacuum where scale separation is broken and the flux vacuum that respects it is also finite, namely
\be
\Delta_{\text{geodesic}}^{\left(S \lesssim -\frac17 \,\to\, S=0\right)} < \Delta_{\Gamma} < +\infty \,.
\ee

\subsubsection{Verifying the swampland distance conjecture}\label{SS-VSDS}

We would now like to find the geodesic distance $\Delta_{\text{geodesic}}$ and estimate the parameter $\gamma$ appearing in the distance criterion relation \eqref{DCformula}. 
Since we know (see Section \ref{Setup}) that, while moving from the non-scale-separated critical configuration to the scale-separated one, the KK modes, e.g. of the dilaton $\phi$, become light as far as the directions associated with the radii $r_2$, $r_4$ and $r_6$ are concerned, it could be worth to test the swampland distance conjecture in such explicit construction. 
We thus refer to the formula 
\be \label{rSDC}
m_{\text{KK}}^{(\phi)}(S=0) = m_{\text{KK}}^{(\phi)} \left( S\lesssim-\frac17 \right) e^{- \gamma \Delta_{\text{geodesic}}} \,. 
\ee

We are now going to present some numerical estimates for $\gamma$ when, 
coherently with the quantization conditions, the flux parameters $h$ and $m$ are fixed to be 
\be
h = (2\pi)^2 16 \quad\quad \text{and} \quad\quad m = \frac{1}{2\pi} \,, 
\ee
the flux $f$ takes the values 
\be
f = (2\pi)^3 N  \quad\quad \text{with} \quad\quad N=10, \, 10^2, \, 10^3, \, 10^5\,, 
\ee
and the flux $q$ takes two different values on the boundaries of the path. 
In particular, on the non-scale-separated side we have $q=(2\pi)^3$ and, as we said above, $f = (2\pi)^3 N$; 
on the scale-separated side, instead, we have $q=f=(2\pi)^3 N$. 
We will estimate the distance for different $N$ values, because, such an increase makes more and more clean the distinction between the broken scale-separated regime and the circumstance where scale separation is realized, 
also by moving the corresponding two vacua away one from the other more and more in field space.

In order to find the geodesic path we have to identify the geometric structure of the scalar field space and specifically show that it is a $\mathbb{H}^2 \times \mathbb{R}^3$ space.\footnote{We would like to thank Vincent Van Hemelryck for pointing this out.} To this end we have to perform some useful field redefinitions: we firstly trade $\theta$ for 
\be
\theta = \tilde \theta - \frac16 \eta
\ee
and then we exchange $\tilde \theta$ and $\eta$ with
\be
\tilde \theta = \frac{\hat \theta}{\sqrt{270}} \quad\quad \text{and} \quad\quad \eta = \sqrt{\frac{2}{21}} \hat \eta \,.
\ee
Moreover, we introduce the new coordinates $\{u_1,u_2,u_3,u_4\}$ via an appropriate rotation:
\be
\left(\begin{matrix} \phi \\ v \\ \hat \theta \\ \hat \eta \end{matrix}\right) = \frac{1344}{533+3024\beta^2} \left(\begin{matrix}  -\frac18 & -\frac{3\beta}{2} & -\frac{\sqrt{\frac{10}{3}}}{3} & -\frac{\sqrt{\frac{2}{21}}}{3} \\ \frac{3\beta}{2} & -\frac18 & -\frac{\sqrt{\frac{2}{21}}}{3} &  \frac{\sqrt{\frac{10}{3}}}{3} \\ \frac{\sqrt{\frac{10}{3}}}{3} &  \frac{\sqrt{\frac{2}{21}}}{3} & -\frac18 & -\frac{3\beta}{2}  \\ \frac{\sqrt{\frac{2}{21}}}{3} & -\frac{\sqrt{\frac{10}{3}}}{3} & \frac{3\beta}{2} & -\frac18 \end{matrix}\right)  \left(\begin{matrix} u_1 \\ u_2 \\ u_3 \\ u_4 \, \end{matrix}\right), 
\ee
where the matrix
\be
O = \left(\begin{matrix}  -\frac18 & -\frac{3\beta}{2} & -\frac{\sqrt{\frac{10}{3}}}{3} & -\frac{\sqrt{\frac{2}{21}}}{3} \\ \frac{3\beta}{2} & -\frac18 & -\frac{\sqrt{\frac{2}{21}}}{3} &  \frac{\sqrt{\frac{10}{3}}}{3} \\ \frac{\sqrt{\frac{10}{3}}}{3} &  \frac{\sqrt{\frac{2}{21}}}{3} & -\frac18 & -\frac{3\beta}{2}  \\ \frac{\sqrt{\frac{2}{21}}}{3} & -\frac{\sqrt{\frac{10}{3}}}{3} & \frac{3\beta}{2} & -\frac18 \end{matrix}\right) \quad\quad \text{with} \quad \text{det}O = \left(\frac{533+3024\beta^2}{1344}\right)^2
\ee
is such that
\be
O O^T = (\text{det}O)^{\frac12} \mathbb{1}_4 \,,
\ee
thus representing an O(4) transformation.
Once the field $\chi$ is also redefined to be
\be
\chi = \frac{1}{2 (\text{det}O)^{\frac14}} h_1 \,,
\ee
the expression \eqref{MasterDelta} for the field space distance takes the form
\be
\Delta = \frac{1}{\sqrt{2} (\text{det}O)^{\frac14}} \int_0^1 \text{d}\xi \sqrt{e^{-2 u_1} \left(\frac{\text{d}h_1}{d\xi}\right)^2 + \left(\frac{\text{d}u_1}{d\xi}\right)^2 + \left(\frac{\text{d}u_2}{d\xi}\right)^2 +\left(\frac{\text{d}u_3}{d\xi}\right)^2 + \left(\frac{\text{d}u_4}{d\xi}\right)^2}
\ee
so that, after trading $u_1$ for $h_2 = e^{-u_1}$, becomes
\be\label{Geodesic1}
\Delta = \frac{1}{\sqrt{2} (\text{det}O)^{\frac14}} \int_0^1 \text{d}\xi \sqrt{\frac{1}{h_2^2} \left[\left(\frac{\text{d}h_1}{d\xi}\right)^2 + \left(\frac{\text{d}h_2}{d\xi}\right)^2 \right] + \left(\frac{\text{d}u_2}{d\xi}\right)^2 +\left(\frac{\text{d}u_3}{d\xi}\right)^2 + \left(\frac{\text{d}u_4}{d\xi}\right)^2}
\ee
and makes evident the $\mathbb{H}^2 \times \mathbb{R}^3$ structure that we have mentioned above.
The geodesic path is then \cite{Shiu:2022oti}
\be
h_1 = l \sin\left[f(\xi)\right] + h_{1,0} \,, \quad\quad h_2 = l \cos\left[f(\xi)\right] \,,
\ee
with
\be
f(\xi) = 2\arctan\left[\tanh\left[\frac{d_1 \xi + d_2}{2}\right]\right] \,,
\ee
and
\be
u_2 = d_3 \xi + d_4 \,, \quad\quad u_3 = d_5 \xi + d_6 \,, \quad\quad u_4 = d_7 \xi + d_8 \,.
\ee
On the geodesic path just specified, the geodesic distance \eqref{Geodesic1} evaluates to
\be\label{Geodesic2}
\Delta = \frac{1}{\sqrt{2} (\text{det}O)^{\frac14}}  \sqrt{d_1^2 + d_3^2 + d_5^2 + d_7^2} \,,
\ee
where the coefficients $\{d_i\}_{i=1,3,5,7}$ are determined by the boundary values of the fields $h_1$, $h_2$, $u_2$, $u_3$ and $u_4$ (respectively) corresponding to the two vacuum configurations we are interpolating between: the non-scale-separated one at $\xi=0$ and the scale-separated one at $\xi=1$\footnote{Taking into account the various field redefinitions, the boundary values of $h_1$, $h_2$, $u_2$, $u_3$ and $u_4$ are extracted from those of $\chi$, for which we assume $\chi(\xi=0) \sim 1$ and $\chi(\xi=1) \sim N$, $\phi$, $v$, $\theta$ and $\eta$, which come from the extremization procedure. As we also pointed out earlier, 
the $\chi$ represents a collective movement of $N$ (redefined) scalars, 
that is why we take its boundary value at $\xi=1$ to be of order $N$.}.

When $f = (2\pi)^3 10$, the parameters the geodesic distance \eqref{Geodesic2} depends on take the values
\be
d_1 = 1.563411 \,, \quad\quad d_3 = -0.031082 \,, \quad\quad d_5 = 0.126126 \,, \quad\quad d_7 = 0.395854 
\ee
and, as a consequence, $\gamma = 0.0927954$.
For $f = (2\pi)^3 10^2$, we find
\be
d_1 = 3.124955 \,, \quad\quad d_3 = -0.062164 \,, \quad\quad d_5 = 0.252252 \,, \quad\quad d_7 = 0.791708 \,,
\ee
so that $\gamma = 0.0928472$.
If $f = (2\pi)^3 10^3$, the relevant parameters are
\be
d_1 = 4.687406 \,, \quad\quad d_3 = -0.093246 \,, \quad\quad d_5 = 0.378378 \,, \quad\quad d_7 = 1.18756  \,,
\ee
and we are able to estimate $\gamma$ as $\gamma = 0.0928477$.
As $f = (2\pi)^3 10^5$, we have
\be
d_1 = 5.5097572 \,, \quad\quad d_3 = -0.15541 \,, \quad\quad d_5 = 0.63063 \,, \quad\quad d_7 = 1.97927 \,,
\ee
and the distance criterion parameter takes the value $\gamma = 0.127445$.

As the reader can appreciate, the calculation of the geodesic distance \eqref{Geodesic2} allows to constrain the parameter $\gamma$ to be roughly a $\mathcal{O}(1)$ number, coherently with what the distance conjecture prescribes.

\section{Discussion}

In this work we have studied a flux compactification from ten to three dimensions of massive IIA supergravity with orientifold planes. 
We have found that within the same setup, and just by changing a single flux number, we can have access to a background that can or can not have scale separation. 
The setup where scale separation is broken corresponds to an external AdS$_3$ with 3 internal dimensions comparable to the AdS length scale while the other four remain parametrically smaller. 
We have analyzed the masses of the closed string moduli and we have studied their properties, for instance extracting the dimensions of the corresponding putative CFT dual operators. 
We have also seen that the use of an open string modulus related to a D4-brane (or a stack of D4-branes) allows to interpolate between the broken scale-separated configuration and the scale-separated one. Moreover, while moving from the latter regime to the former one we have been able to explicitly check, via the dilaton KK modes, the swampland distance conjecture. 

A few directions that can be pursued further are the following.
It is clearly worth to explicitly study the properties of these AdS$_3$ vacua, but also the earlier constructions \cite{Farakos:2020phe,VanHemelryck:2022ynr}, with respect to the O-plane backreaction and verify the open string moduli stabilization. 
Since the internal space that we have used here is an orbifold, one could also wonder whether the internal space should have a smooth description without singularities. 
Clearly, another open question is to find the most general flux ansatz for the system that we have studied here (and in \cite{Farakos:2020phe}) which still allows for moduli stabilization, and subsequently check when scale separation can be achieved. It is possible that other interesting flux choices can be found with other interesting properties. 
Finally one can only wonder if a similar effect, i.e. the option to turn scale separation on and off, can also be done in four dimensions in \cite{DeWolfe:2005uu}.

\section*{Acknowledgements} 

We thank Maxim Emelin and Thomas Van Riet for discussions and we especially thank Vincent Van Hemelryck for discussions, comments on the first draft and helpful suggestions on section 4. 
We would also like to thank Leopoldo Cerbaro for his help in the realization of the plots. 
The work of F.F. is supported by the MIUR-PRIN contract 2017CC72MK003. 
F. F. would like to thank the Laboratoire d'Annecy-le-Vieux de Physique Th\'eorique (LAPTh) for the hospitality during the early stages of this work.

\appendix

\section{Flux quantization}\label{FluxAppendix}

For the ten-dimensional theory under consideration, after setting $\alpha^\prime = 1$, the quantization rules are
\be
\int F_p = (2\pi)^{p-1} f_p \quad\quad \text{with } \ f_p \in \mathbb{Z}
\ee
and
\be
\text{d}F_{8-p} = \dots + (2\pi)^7 Q_{\rm{ Dp/Op}} \delta_{9-p} \,.
\ee
Moreover, the charge of a single Dp-brane in $\alpha^\prime$ units is
\be
Q_{\rm Dp} = \frac{1}{(2\pi)^p} 
\ee
and the charge of an Op-plane is then
\be \label{Opcharge}
Q_{\rm Op} = - 2^{p-5} Q_{\rm Dp} \,.
\ee
If we denote by $K$ and $M$ the quantum numbers associated with the $H_3$ flux and the Romans mass, respectively, the integrated Bianchi identity for the $F_2$ flux gives the RR tadpole cancellation condition for each cycle, i.e.
\be
K M = 2 N_{\rm O6} - N_{\rm D6} \,,
\ee
where $N_{\rm D6}$ and $N_{\rm O6}$ represent the numbers of D6-branes and O6-planes; the factor of 2 accompanying $N_{\rm O6}$ is due to \eqref{Opcharge}, $Q_{\rm O6}$ being $Q_{\rm O6} = - 2 Q_{\rm D6}$. Since each O6 involution has eight fixed points (corresponding to the points $\{0, \frac12\}$) in the covering space, when no D6-branes are in the game, for an AdS vacuum we have
\be
h = (2\pi)^2 K \,, \quad m = (2\pi)^{-1} M \,, \quad f = (2\pi)^3 N \,,
\ee
where $K$, $M$, $N \in \mathbb{Z}$.

\section{Estimating the Chern--Simons term scaling} 
\label{ChernSimons}

In this appendix we would like to evaluate the scaling of the Chern--Simons piece of the D4-brane action, in order to justify (even with this crude analysis) why we have focused on the DBI part of $S_{\rm D4}$ in the main part of the paper. The expectation that the Chern--Simons contribution to the D4-brane action does not change the discussion was already pointed out in \cite{Shiu:2022oti}. 
The standard general expression for the bosonic part of the Chern--Simons term of $S_{\text{D4}}$ is up to higher curvature terms 
\begin{align}
    S_{\text{CS}} = T_{p} \int_{p+1} P\Big[\sum_q C_q\Big]\wedge e^{2\pi\alpha^{\prime}F-B_2} \,.
\end{align}
Since, as we discussed in the bulk of the paper, $B_2 \wedge C_3$ contributions do not appear with legs on the D4-brane world-volume, the Chern--Simons term of interest is 
\be\label{CSterm}
T_4 \int_{\text{AdS}_3 \times \Sigma_2} C_5 \,, 
\ee
where $\Sigma_2$ is the 2-cycle wrapped by the D4-brane. 
To find the scaling of $C_5$ we can use the Einstein frame relation between the magnetic and the electric fluxes, namely 
\be
F_6 = e^{\phi/2} \star_{10} F_4 \,. 
\ee
Since the only $C_5$ RR field that can be space-filling with respect to the D4-brane world-volume is the one that is Poincar\'e-dual to the $F_{4,q}$ flux, we simply need to dualize that. 
The Poincar\'e dual of a $p$-form in D dimensions is given by 
\begin{align}
\label{HodgeDual}
\star_D F_p\sim\sqrt{G_{D}}f_{a_1\dots a_p} g^{a_1b_1}\dots g^{a_pb_p}\epsilon_{b_1\dots b_p b_{p+1}\dots b_{d}}\text{d}y^{b_{p+1}\dots b_{d}} \,, 
\end{align}
where we stress that one should use the metric ansatz \eqref{metricansatz}. 
Then, we have 
\be
\begin{aligned}
C_{5,q} &\sim F_{6,q} \sim e^{\phi/2} \star_{10} F_{4,q} \sim e^{\phi/2} \, \sqrt{-G} \, (G_{11} G_{33} G_{55} G_{77})^{-1} \, q 
\\[1mm]
&\sim  N^{-\frac{3 + 7S}{8}} \times N^{-\frac{21}{16} (7+11S)} \times N^{\frac{7}{16} (7 + 11S)} \times N^{-12 S - \frac{7+11S}{2}} \times N^{1 + 7S} \sim N^{-9-21S}
\,. 
\end{aligned}
\ee
We see that the Chern--Simons term \eqref{CSterm} scales with $N$ in such a way that its contribution is not big. Moreover, the $\psi$ dependence is not manifest in the previous formula, because it is essentially hidden in the internal space radii.
As a result, we find that $V_{\rm CS}/|V_{\rm flux}| \sim N^{-1 -7S}$, which is twice the damping of the DBI piece in $S_{\rm D4}$. In particular, when $S=0$, the ratio $V_{\rm CS}/|V_{\rm flux}|$ scales as $N^{-1}$ and, when $S=-1/7$, it scales like $N^0$.


\end{document}